%% file: sample.tex
\documentclass[sigconf]{acmart}
\usepackage{balance}
\usepackage{bm}
\usepackage{threeparttable}
\usepackage{booktabs} 
\usepackage{multirow}
\usepackage{subfigure}
\usepackage{enumitem}
\usepackage{balance}
\usepackage{diagbox}
\usepackage{enumitem}
\setlist[itemize]{leftmargin=*}

\usepackage{amsmath} 
 
\usepackage{amssymb}
\usepackage{pdfrender}

\usepackage[hang,flushmargin]{footmisc}

\copyrightyear{2023}
\acmYear{2023}
\setcopyright{rightsretained}
\acmConference[WWW '23]{Proceedings of the ACM Web Conference 2023}{May 1--5, 2023}{Austin, TX, USA}
\acmBooktitle{Proceedings of the ACM Web Conference 2023 (WWW '23), May 1--5, 2023, Austin, TX, USA}
\acmDOI{10.1145/3543507.3583312}
\acmISBN{978-1-4503-9416-1/23/04}

\usepackage{etoolbox}
\makeatletter
\patchcmd{\maketitle}{\@copyrightpermission}{
  \begin{minipage}{0.3\columnwidth}
     \href{https://creativecommons.org/licenses/by/4.0/}{\includegraphics[width=0.90\textwidth]{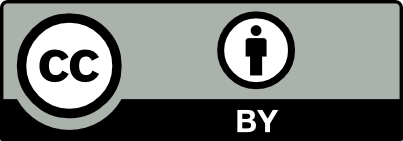}}
  \end{minipage}\hfill
  \begin{minipage}{0.7\columnwidth}
     \href{https://creativecommons.org/licenses/by/4.0/}{This work is licensed under a Creative Commons Attribution International 4.0 License.}
  \end{minipage}
  
  \vspace{5pt}
}{}{}

\makeatother

\begin{document}
\title{Compressed Interaction Graph based Framework for Multi-behavior Recommendation}



\author{Wei Guo}
\email{guowei67@huawei.com}
\authornote{Both authors contributed equally to this research.}
\affiliation{
  \institution{Huawei Noah's Ark Lab}
  \city{Shenzhen}
  \country{China}
}

\author{Chang Meng}
\email{mengc21@mails.tsinghua.edu.cn}
\authornotemark[1]
\authornote{Work done when they were research interns at Huawei Noah’s Ark Lab.}
\affiliation{
  \institution{Tsinghua Shenzhen International Graduate School, Tsinghua University}
  \city{Shenzhen}
  \country{China}}

\author{Enming Yuan}
\email{yem19@mails.tsinghua.edu.cn}
\authornotemark[2]
\affiliation{
  \institution{Institute for Interdisciplinary Information Sciences, Tsinghua University}
  \city{Beijing}
  \country{China}
}

\author{Zhicheng He}
\email{hezhicheng9@huawei.com}
\affiliation{
  \institution{Huawei Noah's Ark Lab}
  \city{Shenzhen}
  \country{China}
}

\author{Huifeng Guo}
\email{huifeng.guo@huawei.com}
\affiliation{
  \institution{Huawei Noah's Ark Lab}
  \city{Shenzhen}
  \country{China}
}

\author{Yingxue Zhang}
\email{yingxue.zhang@huawei.com}
\affiliation{
  \institution{Huawei Technologies Canada}
  \city{Montreal}
  \country{Canada}
}

\author{Bo Chen}
\email{chenbo116@huawei.com}
\affiliation{
  \institution{Huawei Noah's Ark Lab}
  \city{Shenzhen}
  \country{China}
}

\author{Yaochen Hu}
\email{yaochen.hu@huawei.com}
\affiliation{
  \institution{Huawei Technologies Canada}
  \city{Montreal}
  \country{Canada}
}

\author{Ruiming Tang}
\email{tangruiming@huawei.com}
\authornote{Corresponding author.}
\affiliation{
  \institution{Huawei Noah's Ark Lab}
  \city{Shenzhen}
  \country{China}}

\author{Xiu Li}
\email{li.xiu@sz.tsinghua.edu.cn}
\authornotemark[3]
\affiliation{
  \institution{Tsinghua Shenzhen International Graduate School, Tsinghua University}
  \city{Shenzhen}
  \country{China}}

\author{Rui Zhang}
\email{rayteam@yeah.net}
\affiliation{
  \institution{ruizhang.info}
  \city{Shenzhen}
  \country{China}
}

\renewcommand{\shortauthors}{Wei Guo and Chang Meng, et al.}

\begin{abstract}
Multi-types of user behavior data (e.g., clicking, adding to cart, and purchasing) are recorded in most real-world recommendation scenarios, which can help to learn users’ multi-faceted preferences.
However, it is challenging to explore multi-behavior data due to the unbalanced data distribution and sparse target behavior, which lead to the inadequate modeling of high-order relations when treating multi-behavior data ``\textsl{as features}'' and gradient conflict in multi-task learning when treating multi-behavior data ``\textsl{as labels}''.
In this paper, we propose CIGF, a Compressed Interaction Graph based Framework, to overcome the above limitations. 
Specifically, we design a novel Compressed Interaction Graph Convolution Network (CIGCN) to model \textbf{\textsl{instance-level}} high-order relations explicitly.
To alleviate the potential gradient conflict when treating multi-behavior data ``as labels'', we propose a Multi-Expert with Separate Input (MESI) network with \textbf{\textsl{separate input}} on the top of CIGCN for multi-task learning.
Comprehensive experiments on three large-scale real-world datasets demonstrate the superiority of CIGF. 
Ablation studies and in-depth analysis further validate the effectiveness of our proposed model in capturing high-order relations and alleviating gradient conflict.
The source code and datasets are available at \url{https://github.com/MC-CV/CIGF}.


\noindent\let\thefootnote\relax\footnotetext{$\ast$ indicates co-first authors with equal contributions. \\ $\dagger$ Work done when they were research interns at Huawei Noah’s Ark Lab. \\ $\ddagger$ indicates the co-corresponding authors.}

\end{abstract}

\ccsdesc[500]{Information systems~Recommender systems}

\keywords{Multi-behavior Recommendation, Interaction Graph, Multi-task}
\maketitle

\input{sections/introduction.tex}
\input{sections/related_work.tex}
\input{sections/preliminary.tex}

\input{sections/method.tex}

\input{sections/experiments.tex}
\input{sections/conclusion.tex}
\balance


\begin{acks}

This work was partly supported by the Science and Technology Innovation 2030-Key Project under Grant 2021ZD0201404 and Aminer·
ShenZhen·ScientificSuperBrain. 
And we thank MindSpore \cite{mindspore} for the partial support of this work, which is a new deep learning computing framework.

\end{acks}

\bibliographystyle{ACM-Reference-Format}
\bibliography{sample}
\input{appendix.tex}

\end{document}

%% file: sections/introduction.tex
\begin{figure}[H]
	\centering
	\setlength{\belowcaptionskip}{0cm}
	\setlength{\abovecaptionskip}{0cm}
	\includegraphics[width=0.45\textwidth]{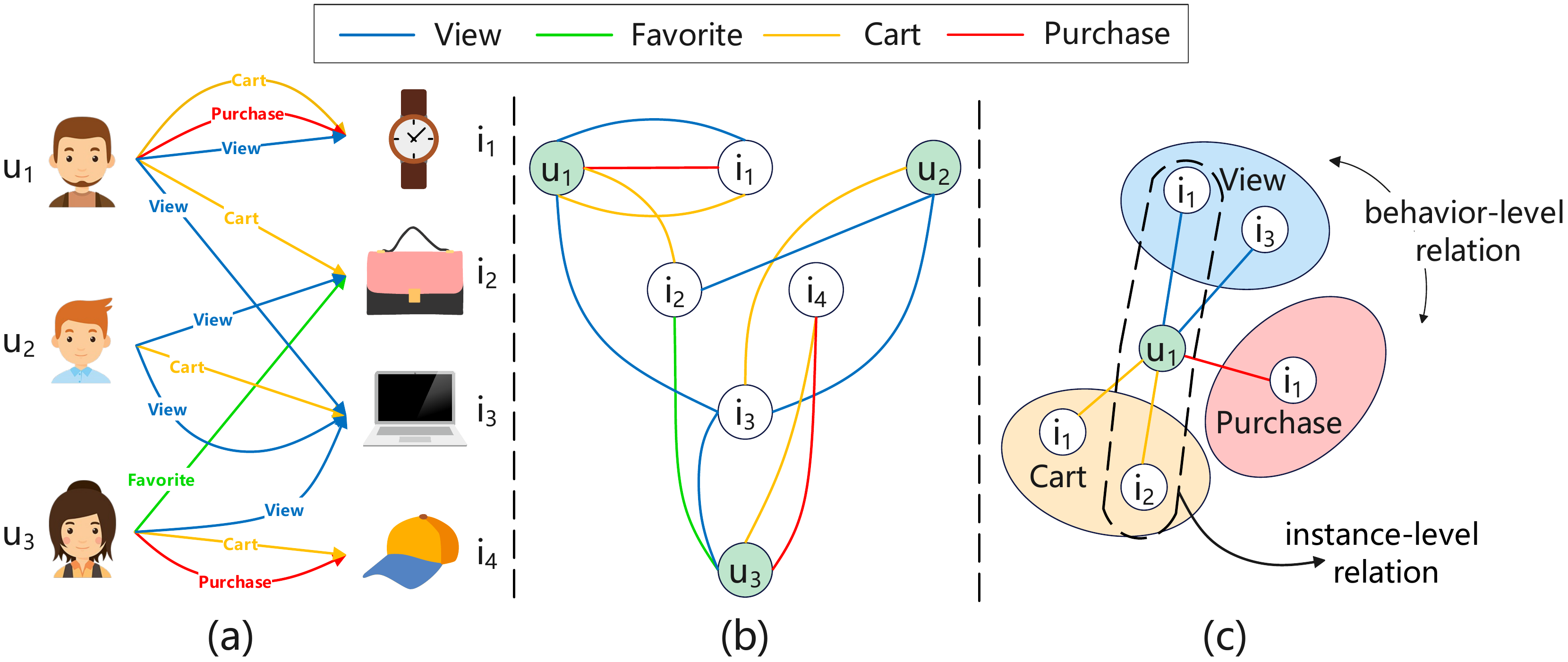}
    \Description{An example of multiple types of behaviors on an e-commerce website, the corresponding MBG and a comparison of behavior-level and instance-level relation.}
	\caption{An example of multiple types of behaviors on an e-commerce website, the corresponding MBG and a comparison of behavior-level and instance-level relation. }
	\label{fig:relation}
	\vspace{-3mm}
\end{figure}
\section{Introduction}\label{introduction}
Recommender systems (RS) serve as an important tool to meet personalized information needs.
To predict users' preferences for items, various methods have been devoted to \textsl{Collaborative Filtering} (CF) \cite{cfsurvey} techniques, which learn user and item representations from their historical interactions and then make predictions based on these representations. 
Most CF methods \cite{bpr,deepmf,ncf,ngcf,lightgcn} are designed for a single type of behavior and rarely consider users' multi-faceted preferences, which widely exist in real-world web applications.
Take the example of an e-commerce website, as shown in Figure 1(a).  Users interact with items through different behaviors, such as viewing, adding to cart, tagging as favorites, and purchasing. 
Since different types of behaviors exhibit different interactive patterns out of users' diverse interests, it's of great importance to explicitly leverage multi-behavior data for recommendation. 

NMTR \cite{nmtr}, DIPN \cite{dipn}, and MATN \cite{matn} regard multiple behaviors as different types and employ neural collaborative filtering unit, attention operator, and transformer architecture to model their dependencies, which perform much better than treating them as the same type. 
Multi-behavior data can be regarded as a multiplex bipartite graph (MBG), as shown in Figure 1(b).
Recently, thanks to its capacity in representing relational information and modeling high-order relations which carry collaborative signals among users and items, graph neural networks (GNNs) based models \cite{ngcf,lightgcn,lr-gccf,meng2022coarse} have become popular for recommendation.
For example, MBGCN \cite{mbgcn}, GHCF \cite{ghcf}, and MB-GMN \cite{mbgmn} further empower GNNs with multi-graph, non-sampling, and meta network to capture high-order collaborative signals on multiplex bipartite graphs.





\begin{figure}[!t]
	\setlength{\belowcaptionskip}{-0cm}
	\setlength{\abovecaptionskip}{-0.1cm}
	\subfigure{
        \begin{minipage}[t]{0.3\linewidth}
        \centering
		\label{fig:distribution_beibei} 
		\includegraphics[width=1.0in]{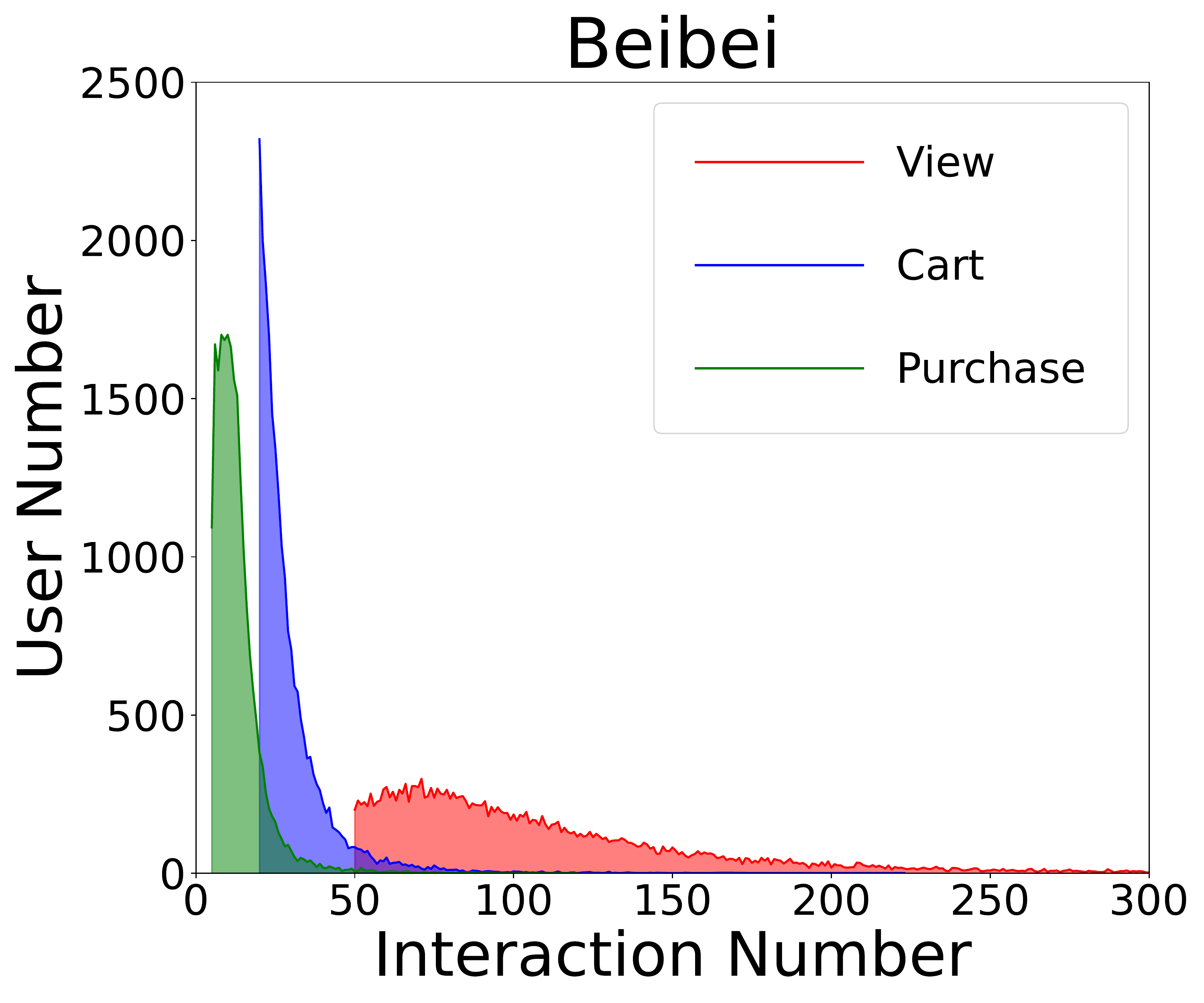}
        \end{minipage}}
	\subfigure{
        \begin{minipage}[t]{0.3\linewidth}
        \centering
		\label{fig:distribution_taobao} 
		\includegraphics[width=1.0in]{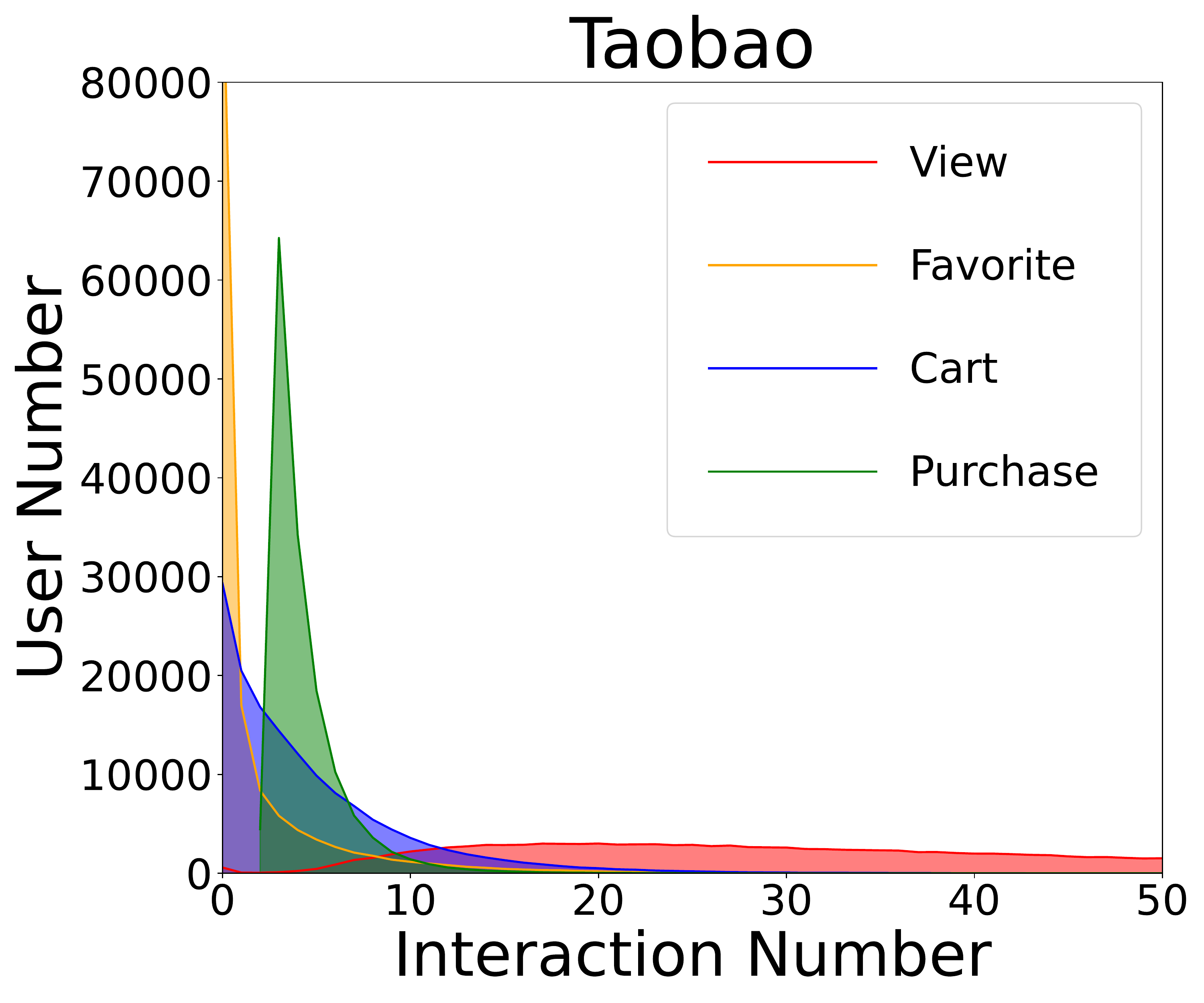}
        \end{minipage}}
    \subfigure{
        \begin{minipage}[t]{0.3\linewidth}
        \centering
		\label{fig:distribution_taobao} 
		\includegraphics[width=1.0in]{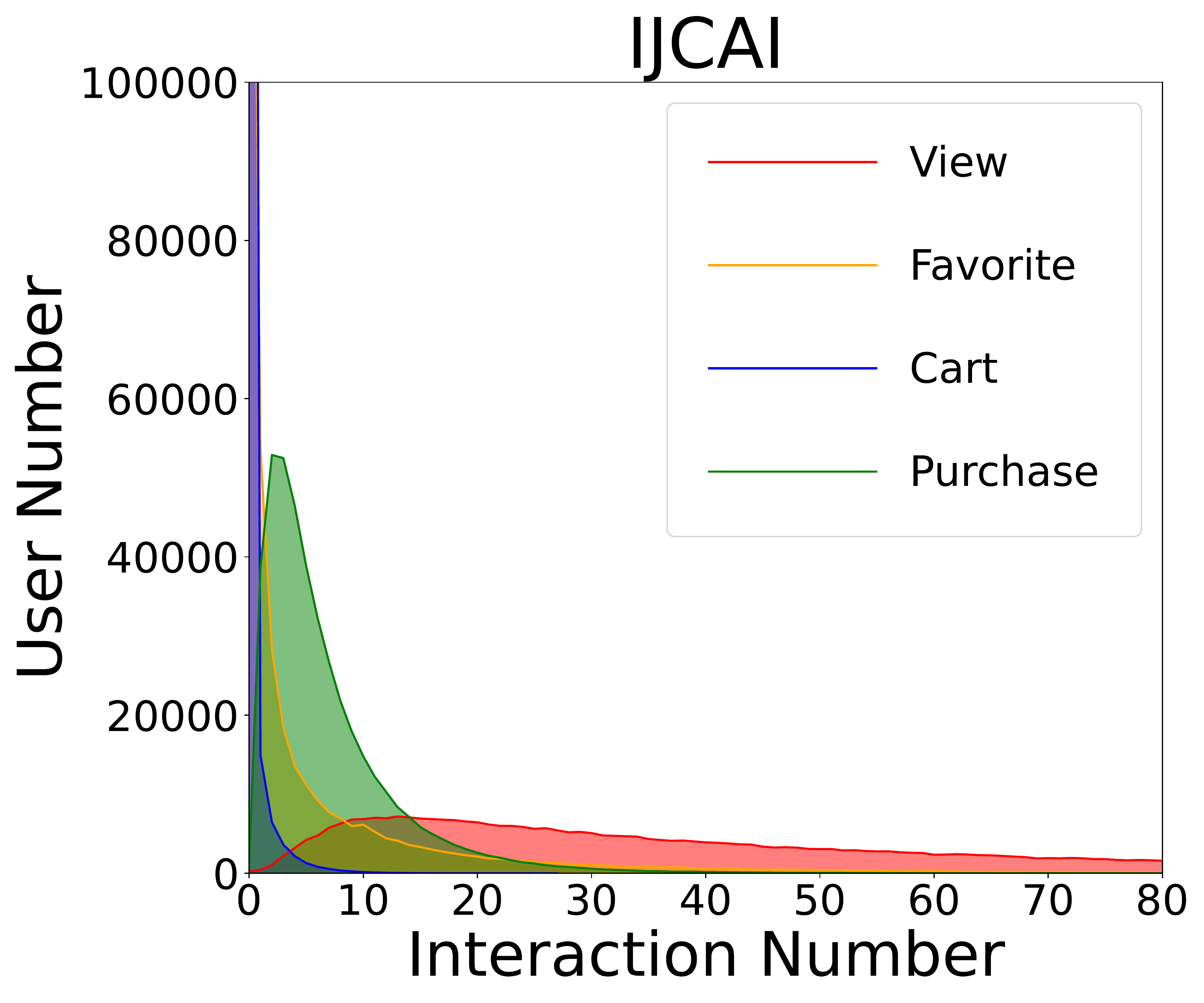}
        \end{minipage}}
    \Description{Histogram of user numbers w.r.t interaction numbers for different behaviors.}
	\caption{Histogram of user numbers w.r.t interaction numbers for different behaviors.}
	\vspace{-3mm}
	\label{fig:distribution}
\end{figure}

Multi-behavior data can be treated ``\textsl{as features}'' for multi-behavior relation learning or ``\textsl{as labels}'' for multi-task supervised learning.
Despite years of research, two challenges remain:
\begin{itemize}
\item \textbf{Unbalanced Data Distribution.}
As we can see from Figure \ref{fig:distribution}, observed interactions are highly unbalanced for different users and different behaviors, where a small percentage of users and behaviors cover most of the interactions. 
\item \textbf{Sparse Target Behavior} (behavior to be predicted, e.g., purchase in e-commerce).
We can also find that most users have less than 10 purchase records, which is extremely sparse compared with the whole item space with thousands to millions of items.
\end{itemize}
We dig into these challenges and observe the following limitations: 

\begin{itemize}
\item \textbf{Inadequate modeling of high-order relations when treating  multi-behavior data ``\textsl{as features}''.}
User-item relations are meaningful for revealing the underlying reasons that motivate users' preference on items.
For example, as shown in Figure 1(a), there are several third order relations between $u_1$ and $i_4$ (e.g., $u_1 \xrightarrow{cart} i_2 \xrightarrow{be \; favored \; by} u_{3} \xrightarrow{purchase} i_{4}$).
With the help of collaborative effect, we predict that $u_1$ is likely to purchase $i_4$ as $u_3$, the user similar to $u_1$, has purchased $i_4$ before.
Existing methods like MBGCN, GHCF, and MB-GMN have attempted to employ GNNs to incorporate high-order relations. 
However, they use a two-stage paradigm which first learns representation for each behavior by considering all historical records belonging to this behavior, then leveraging the learned representation to model high-order relation across different behaviors.
We argue that this relation modeling manner is \textbf{\textsl{behavior-level}}, as depicted in Figure \ref{fig:relation}(c).
Due to the unbalanced data distribution, the learned relations are easily biased toward high-degree users and behaviors, and thus making the learned representations unable to effectively capture high-order relations.


\item \textbf{Potential gradient conflict when treating multi-behavior data ``\textsl{as labels}''.}
Early works like MBGCN \cite{mbgcn} and MATN \cite{matn} only use target behavior as labels to train the model, which is vulnerable to the sparsity problem due to the sparse target behavior.
To alleviate this problem, it is promising to use auxiliary behaviors as labels with multi-task learning (MTL) techniques.
However, it is not easy to train with multiple objectives due to the \textsl{negative transfer}\footnote{We ignore the \textsl{seesaw phenomenon} (i.e., MTL models improve the performances of some tasks while sacrifices the others) \cite{PLE} here as the objective is to predict the target behavior in the multi-behavior recommendation.} \cite{negativeTransfer} phenomenon.
\textsl{Negative transfer} indicates the performance deterioration when knowledge is transferred across different tasks. 
Therefore, it's risky to treat multi-behavior data "\textsl{as labels}".
Several recent works like NMTR \cite{nmtr}, GHCF \cite{ghcf}, and MB-GMN \cite{mbgmn} have investigated MTL in multi-behavior recommendation.
As they use the \textbf{\textsl{same input}}, these methods might suffer from the gradient conflict due to the coupled gradient issue.
Detailed explanations are presented in Section \ref{limitations}.



\end{itemize}

To tackle the above limitations, we propose a novel Compressed Interaction Graph based Framework (CIGF) for better representation learning of users and items.
To handle the inadequate modeling of high-order relations when treating multi-behavior data "\textsl{as features}", we design a Compressed Interaction Graph Convolution Network (CIGCN) to model high-order relations explicitly.
CIGCN firstly leverages matrix multiplication as the interaction operator to generate high-order interaction graphs which encode \textbf{\textsl{instance-level}} high-order relations (including user-user, user-item, and item-item) explicitly, then leverages node-wise attention mechanism to select the most useful high-order interaction graphs and compress the graph space.
Finally, state-of-the-art GCN models are combined with residual connections \cite{resnet} on these graphs to explore high-order graph information and alleviate the over-smoothing issue for representation learning.

To alleviate the potential gradient conflict when treating multi-behavior data "\textsl{as labels}", we propose a Multi-Expert with Separate Input (MESI) network on the top of CIGCN for MTL.
MESI network is a hierarchical neural architecture similar to the MMOE \cite{mmoe} and PLE \cite{PLE}.
However, \textbf{\textsl{separate inputs}} are introduced to replace the \textbf{\textsl{same input}} used in the original MMOE and PLE models for MTL.
Specifically, we use relations starting from different types of behaviors for the learning of \textbf{\textsl{separate inputs}}.
By using \textbf{\textsl{separate inputs}} explicitly to learn task-aware information, potential gradient conflict of the \textbf{\textsl{same input}} can be alleviated when knowledge is transferred across different tasks, which makes the learning process more stable and effective.
Explanations for the decoupled gradient of MESI can be referred to Section \ref{limitations}.

To summarize, our work makes the following contributions:
\begin{itemize}
	\item  We look at the multi-behavior recommendation problem from a new perspective, which treats multi-behavior data "\textsl{as features}" and "\textsl{as labels}" with data analysis and theoretical support.
	\item We propse a novel compressed Interaction Graph based Framework (CIGF) which is composed of a Compressed Interaction Graph Convolution Network (CIGCN) and a Multi-Expert with Separate Input (MESI) network. CIGCN is designed for \textbf{\textsl{instance-level}} high-order relation modeling with explicit graph interaction when treating multi-behavior data "\textsl{as features}".
	MESI is designed to alleviate the potential gradient conflict with \textbf{\textsl{separate inputs}} when treating multi-behavior data "\textsl{as labels}".
	\item We conduct extensive experiments on three real-world datasets to demonstrate the effectiveness of our proposed CIGF framework. 
	The ablation analysis and in-depth analysis further verify the effectiveness and rationality of CIGCN and MESI. Besides, we further analyze the complexity of our method and conduct detailed efficiency experiments in Appendix \ref{complexity and efficiency}.
\end{itemize} 

%% file: sections/related_work.tex
\section{Related Work}\label{related_work}

\begin{figure*}[t]
	\centering
	\setlength{\belowcaptionskip}{-0.0cm}
	\setlength{\abovecaptionskip}{-0.0cm}
	\includegraphics[width=0.95\textwidth]{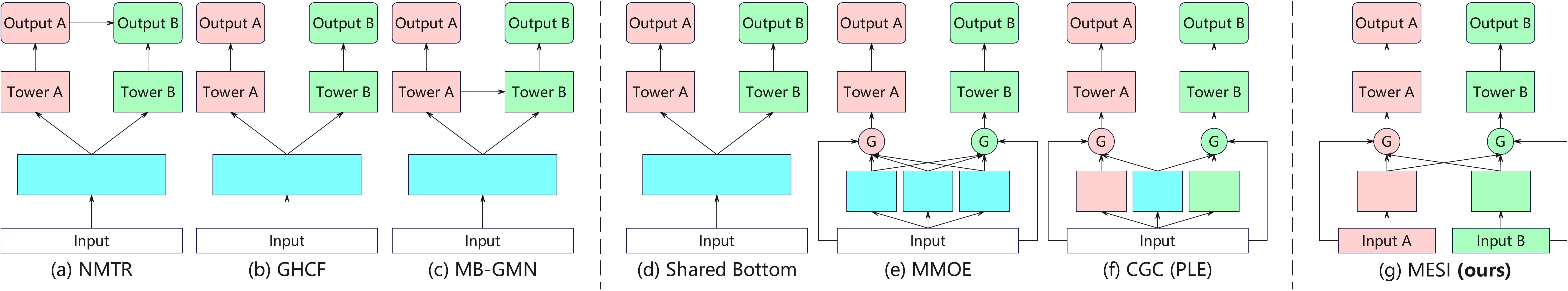}
    \Description{Network structure of existing models and our proposed MESI model. Blue rectangles represent shared layers, pink and green rectangles represent task-specific layers, and pink and green circles denote task-specific gates.}
	\caption{Network structure of existing models and our proposed MESI model. Blue rectangles represent shared layers, pink and green rectangles represent task-specific layers, and pink and green circles denote task-specific gates.}
	\label{fig:multi-behavior learning compare}
	\vspace{-3mm}
\end{figure*}
\textbf{GNNs for Recommendation.}
GNNs based methods can be used for multi-behavior data by treating it ``\textsl{as features}''.
Most of the existing GNNs are proposed for homogeneous graphs, such as NGCF \cite{ngcf}, LR-GCCF \cite{lr-gccf}, and LightGCN \cite{lightgcn}, which ignore the multiple types of edges.
Recently, some researchers have focused on the heterogeneous graph and proposed methods like HGNN \cite{hgnn}, R-GCN \cite{RGNN}, and HGAT \cite{hgat}.
However, these methods merely consider the \textbf{\textsl{behavior-level}} relations by utilizing the behavior-level representations for relation modeling.
Hyper-graph based methods \cite{hypergnn,benson2016higher} leverage hyper-graph to model complex high-order
relations.
However, as an edge in hyper-graph connects two or more nodes, it is not suitable for the multi-behavior case where a node pair connects multiple edges.
Existing meta-path based methods, like Metapath2vec \cite{metapath2vec}, MCRec \cite{mcrec}, and HAN \cite{han} model high-order relations with the manually selected meta-paths, which is limited by the need of expert knowledge and the difficulty of searching all useful meta-paths with arbitrary length and edge types. 
\newline
\textbf{MTL for Recommendation.}
MTL methods can be used for multi-behavior data by treating it ``\textsl{as labels}''.
A widely used model is the shared bottom structure in Figure \ref{fig:multi-behavior learning compare}(d).
Though useful for knowledge sharing within multiple tasks, it still suffers from the risk of conflicts due to the task differences.
To handle the task difference, some studies apply the attention network for information fusion.
MMOE \cite{mmoe} in Figure \ref{fig:multi-behavior learning compare}(e) extends MOE \cite{moe} to utilize different gating networks to obtain different fusion weights in MTL. 
PLE \cite{PLE} in Figure \ref{fig:multi-behavior learning compare}(f) further proposes to leverage shared or task-specific experts at the bottom and then employs gating networks to combine these experts adaptively, thus to handle task conflicts and alleviate the \textsl{negative transfer} issue.
However, they still utilize the \textbf{\textsl{same input}} for MTL.
We argue that this manner might suffer from the gradient conflict due to the coupled gradient issue.
Detailed explanations are presented in Section \ref{limitations}.
\newline
\textbf{Multi-behavior Recommendation.}
\label{Multi-behavior_Recommendation}
Existing multi-behavior recommendation methods can be classified into two categories: graph-based and MTL based \cite{huang2021recent}.
The former category treats multi-behavior data ``\textsl{as features}''.
Some early works like DIPN \cite{dipn} and MATN \cite{matn} fail to capture high-order relations, and thus performing poor.
Most recent works (e.g., GHCF \cite{ghcf} and MBGCN \cite{mbgcn}) use a \textbf{\textsl{behavior-level}} modeling manner that cannot capture the fine-grained \textbf{\textsl{instance-level}} multi-behavior relations.
Some other methods like MBGCN \cite{mbgcn} and MGNN \cite{mgnn} learn high-order relations from the MBG directly, which is difficult to mine useful relations extensively due to the unbalanced data distribution.
Different from the above methods, our proposed CIGCN models high-order relation by explicit graph interaction and graph compression, thus can learn relations in the \textbf{\textsl{instance-level}}.
The latter category treats multi-behavior data ``\textsl{as labels}''.
NMTR \cite{nmtr} in Figure \ref{fig:multi-behavior learning compare}(a) assumes that users’ multiple types of behaviors take place in a fixed order, which may be too strong to be appropriate for all users.
GHCF in Figure \ref{fig:multi-behavior learning compare}(b) uses a similar architecture with shared bottom for MTL.
The only difference is that GHCF uses bilinear operation (Please refer to Section \ref{limitations}) as the prediction head, while shared bottom uses neural network.
MB-GMN \cite{mbgmn} in Figure \ref{fig:multi-behavior learning compare}(c) further proposes to use a meta prediction network to capture the complex cross-type behavior dependency for MTL. 
These existing methods optimize multiple tasks with the same static weights for all samples.
The most obvious drawback is that they can easily suffer from the risk of conflicts caused by sample differences, as different samples may pose different preferences for different tasks.
In contrast, our proposed MESI network learns adaptive weights according to the nature of different samples.
Besides, we utilize the \textbf{\textsl{separate input}} to learn task-aware information to alleviate the potential gradient conflict.

%% file: sections/preliminary.tex
\section{Preliminary}

\subsection{Problem Definition}
In this section, we give the formal definition of the multi-behavior recommendation task.
We denote the user set and item set as $\mathcal{U} = \left\{u_1,u_2,...,u_M\right\}$ and $\mathcal{I} = \left\{i_1,i_2,...,i_N\right\}$, respectively.
The user-item interaction matrices  of behaviors as $\mathcal{Y} = \left\{\mathbf{Y}^1,\mathbf{Y}^2,...,\mathbf{Y}^K\right\}$.  
Where $M$, $N$ and $K$ are the number of users, items and behavior types, respectively, 
and $y_{ui}^k = 1$ denotes that user $u$ interacts with item $i$ under behavior $k$, otherwise $y_{ui}^k = 0$.
Generally, there is a target behavior to be optimized (e.g., purchase), which we denote as $\mathbf{Y}^K$, and other behaviors $\left\{\mathbf{Y}^1,\mathbf{Y}^2,...,\mathbf{Y}^{K-1}\right\}$ (e.g., view and tag as favorite) are treated as auxiliary behaviors for assisting the prediction of target behavior.
The goal is to predict the probability that user $u$ will interact with item $i$ under target behavior $K$.

\subsection{Graph and Relation Definition}
As shown in Figure \ref{fig:relation}(b), we denote the Multiplex Bipartite Graph (MBG)  as $\mathcal{G}=(\mathcal{V}, \mathcal{E}, \mathcal{A})$, where $\mathcal{V} = \mathcal{U}\cup\mathcal{I}$ is the node set containing all users and items, $\mathcal{E} = \cup_{r \in \mathcal{R}}\mathcal{E}_r$ is the edge set including all behavior records between users and items.
Here $r$ denotes a specific type of behavior and $\mathcal{R}$ is the set of all possible behavior types.
$\mathcal{A} = \cup_{r \in \mathcal{R}}\mathbf{A}_r$ is the adjacency matrix set with $\mathbf{A}_r$ denoting adjacency matrix of a specific behavior graph $\mathcal{G}_r=(\mathcal{V},\mathcal{E}_r, \mathbf{A}_r)$.
A relation $\mathcal{P}$ is defined as a path in the MBG $\mathcal{G}$ with the form of $v_1 \xrightarrow{r_1} v_2 \xrightarrow{r_2}  \cdots \xrightarrow{r_l}v_{l+1}$.
We denote $r_\mathcal{P} = \{r_1, r_2, \cdots, r_l\}$ as the set of all edge types in this path.
If $l \geq 2$ and $|r_\mathcal{P}| = 1$, we define this path as a high-order single-behavior relation.
If $l \geq 2$ and $|r_\mathcal{P}| \geq 2$, we define this path as a high-order multi-behavior relation.
Node $v_s$ is node $v_t$'s $l$-th order reachable neighbor if there exists a path connecting node $v_s$ and node $v_t$ and the length of this path is $l$.
In a new generated graph $\mathcal{G}_l$, if arbitrary two connected nodes $v_s$ and $v_t$ are $l$-th order reachable in the original MBG $\mathcal{G}$, $\mathcal{G}_l$ is defined as a $l$-th order graph.
We will illustrate the explicitly modeling of high-order multi-behavior relation through high-order graph interaction and convolution in Section \ref{cigcn}.

\subsection{A Coupled Gradient Issue in MTL}
\begin{figure}[!t]
	\centering
	\setlength{\belowcaptionskip}{-0.4cm}
	\setlength{\abovecaptionskip}{0cm}
	\includegraphics[width=0.15\textwidth]{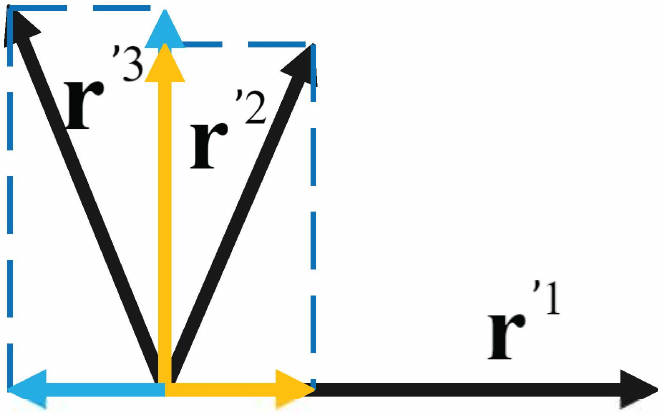}
    \Description{An example of the gradient conflict.}
	\caption{An example of the gradient conflict.}
	\label{fig:array}
	\vspace{-3mm}
\end{figure}
\label{limitations}
Most of the existing methods use the \textbf{\textsl{same input}} for MTL, as summarized in Section \ref{related_work}. 
This may cause a coupled gradient issue in MTL which restricts their learning ability for each task. 
Here we use bilinear module from GHCF \cite{ghcf} as an example to claim this.
The bilinear module can be formulated as:
\begin{equation}
\setlength{\abovedisplayskip}{0.5pt}
\setlength{\belowdisplayskip}{0.5pt}
\hat{o}_{u,i}^{k}={\mathbf{x}_{u}^{*}}^{T} \cdot \operatorname{\textsl{diag}}\left(\mathbf{r}^k\right) \cdot \mathbf{y}_{i}^{*}= \sum_{j}^{d} (\mathbf{x}_{u}^{*} \circ \mathbf{y}_{i}^{*} \circ {\mathbf{r}^k}^{T})_{j}
\end{equation}
where ($\circ$) is the hadamard product operation, $\hat{o}_{u,i}^{k}$ denotes the predictive value of the \textsl{k}-th behavior, $\mathbf{x}_{u}^{*}$ and $\mathbf{y}_{i}^{*}$ represent the learned representation for user $u$ and item $i$.
$\mathbf{r}^k \in \mathbb{R}^{1 \times d}$ is a behavior-aware transformation vector, which projects user and item representation to separate prediction head for MTL, and $d$ denotes the embedding size.
Here we use the square loss as an example for optimization: 
\begin{equation}
\setlength{\abovedisplayskip}{0.5pt}
\setlength{\belowdisplayskip}{0.5pt}
    \mathcal{L}_{u,i}=\sum_{k=1}^{K} (\hat{o}_{u,i}^{k}-{o}_{u,i}^{k})^{2}
\end{equation}
where ${o}_{u,i}^{k}$ is the true label.
Then we have:
\begin{equation}
\setlength{\abovedisplayskip}{0.5pt}
\setlength{\belowdisplayskip}{0.5pt}
{\partial{\mathcal{L}_{u,i}}\over{\partial{(\mathbf{x}_{u}^{*} \circ \mathbf{y}_{i}^{*})}}}
=\sum_{k=1}^{K}{a_{u,i}^{k}{\mathbf{r}^{k}}}=\sum_{k=1}^{K}{\mathbf{r}^{' k}}
\end{equation}
where $a_{u,i}^{k}$ is an scalar, $\mathbf{r}^{' k}$ is the synthetic gradient from the $k$-th behavior, which determines the updating magnitude and direction of the \textbf{\textsl{same input}} vector $\mathbf{x}_{u}^{*} \circ \mathbf{y}_{i}^{*}$.
We can see that the gradients from all behaviors are coupled.
Figure \ref{fig:array} shows an example of $K=3$.
Assuming $\mathbf{r}^{' 1}$ as a reference vector, we do orthogonal decomposition to all the other vectors.
We can find that the components of other vectors are not in the same direction as the reference vector.
This demonstrates the potential gradient conflict brought by the coupled gradient.
The proof for other methods, loss functions and the decoupled gradient of MESI can be referred to Appendix \ref{proof_of_coupled_gradient} and \ref{decoupled_gradient_of_mesi}.

%% file: sections/method.tex
\section{Our Proposed Method}
\begin{figure*}[t]
	\centering
	\setlength{\belowcaptionskip}{-0.0cm}
	\setlength{\abovecaptionskip}{-0.0cm}
	\includegraphics[width=0.95\textwidth]{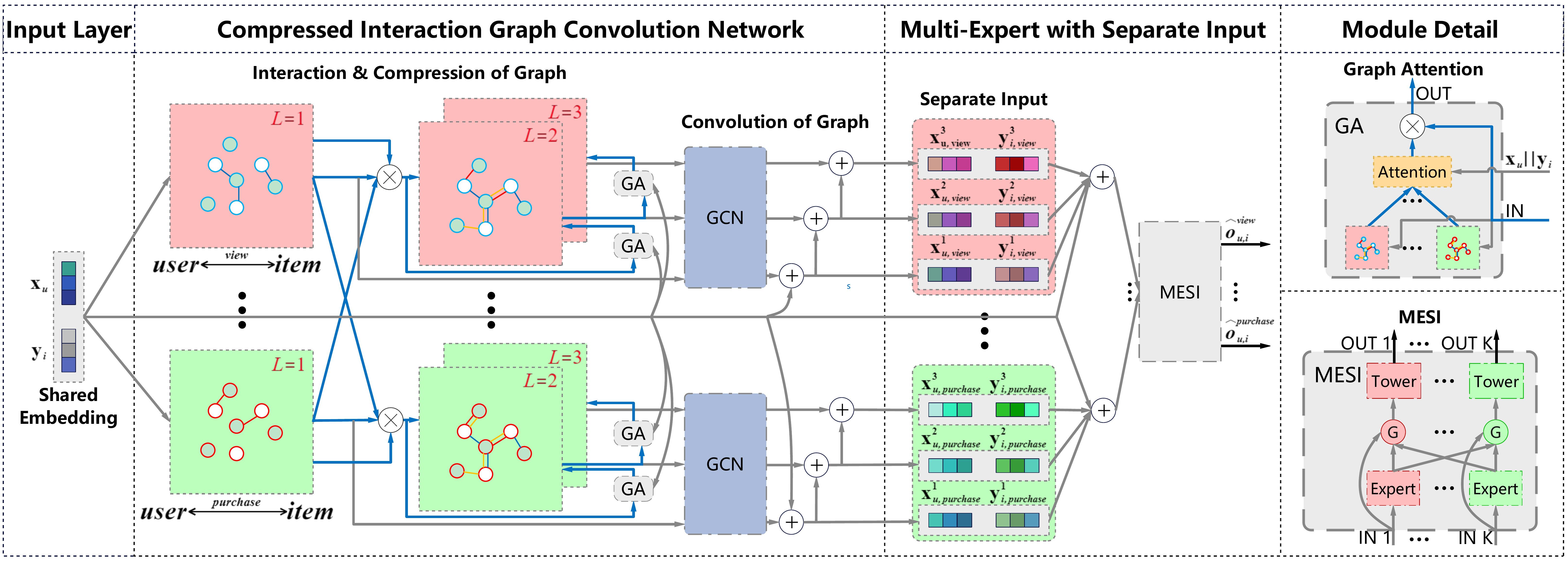}
    \Description{Illustration of the proposed CIGF framework. ($\otimes$) represents the matrix multiplication operation, ($\oplus$) denotes the element-wise addition operation.}
	\caption{Illustration of the proposed CIGF framework. ($\otimes$) represents the matrix multiplication operation, ($\oplus$) denotes the element-wise addition operation.}
	\label{fig:framework}
	\vspace{-3mm}
\end{figure*}

We now present the proposed CIGF framework, which treats multi-behavior data both ``\textsl{as features}'' and ``\textsl{as labels}'' in an end-to-end fashion.
The architecture is shown in Figure~\ref{fig:framework} and it consists of three main components:
i) input layer, which parameterizes users and items as embedding vectors;
ii) compressed interaction graph convolution network (CIGCN), which extracts \textbf{\textsl{instance-level}} high-order relation from the multi-behavior data explicitly by treating it ``\textsl{as features}'';
iii) multi-expert with separate input (MESI) network, which mines multi-task supervision signals from the multi-behavior data with \textbf{\textsl{separate inputs}} by treating it ``\textsl{as labels}''. 

\subsection{Input}
We first apply a shared embedding layer to transform the one-hot IDs of users and items into low-dimensional dense embeddings. 
Formally, given a user-item pair $(u, i)$, the embedding lookup operation for user $u$ and item $i$ can be formulated as follows: 
\begin{equation}
\setlength{\abovedisplayskip}{0.5pt}
\setlength{\belowdisplayskip}{0.5pt}
\mathbf{x}_u = \mathbf{P}^{T} \cdot \mathbf{p}_u, \ 
\mathbf{y}_i = \mathbf{Q}^{T} \cdot \mathbf{q}_i
\end{equation}
where $\mathbf{p}_u \in \mathbb{R}^{M \times 1}$ and $\mathbf{q}_i \in \mathbb{R}^{N \times 1}$ denotes the one-hot IDs of user $u$ and item $i$, $\mathbf{P} \in \mathbb{R}^{M \times d}$ and $\mathbf{Q} \in \mathbb{R}^{N \times d}$ are the user and item embedding matrix and $d$ is the embedding size. 

\subsection{Compressed Interaction Graph
Convolution}
\label{cigcn}
\subsubsection{Graph Interaction}

Inspired by the success of DCN \cite{dcn} and xDeepFM \cite{xdeepfm} which model high-order feature interactions with explicit feature crossing, we use the
adjacency matrix multiplication as the
graph interaction operator for explicit \textbf{\textsl{instance-level}} high-order relation modeling.
As shown in Figure \ref{fig:framework}, we first partition the MBG into several behavior-specified graph $\mathcal{G}^1, \mathcal{G}^2 \cdots \mathcal{G}^K$.
The corresponding adjacency matrices are $\mathbf{A}^1, \mathbf{A}^2 \cdots \mathbf{A}^K$, which can be formulated as:
\begin{equation}
\setlength{\abovedisplayskip}{0.5pt}
\setlength{\belowdisplayskip}{0.5pt}
\mathbf{A}^{k}=\left(\begin{array}{cc}
0 & \mathbf{Y}^{k} \\
\left(\mathbf{Y}^{k}\right)^{T} & 0
\end{array}\right)
\end{equation}
where $\mathbf{Y}^k$ is the user-item interaction matrix of behavior $k$.
We then use these adjacency matrices for explicit high-order graph interaction which encodes the \textbf{\textsl{instance-level}} relations of each two nodes.
Denote the set of all possible $l$-th ($1 \leq l \leq L$) order interaction graph starting from behavior $k$ as $\mathcal{B}_k^l$.
The purpose why we only use interaction graph starting from behavior $k$ here is to generate graph sets with different high-order relations, which will be used as \textbf{\textsl{separate inputs}} for MESI to alleviate the potential gradient conflict.
The generation of $\mathcal{B}_k^l$ can be formulated as:

\begin{equation}
\setlength{\abovedisplayskip}{0.5pt}
\setlength{\belowdisplayskip}{0.5pt}
\mathcal{B}_k^l = \mathcal{B}_k^{l-1} \otimes \{\mathbf{A}^1, \mathbf{A}^2, \cdots,\mathbf{A}^K\} 
\end{equation}
where $\mathcal{B}_k^1 = \{\mathbf{A}^k\}$. ($\otimes$) denotes the matrix multiplication operation between any pairs of matrices from the two sets separately. 
Noticed that there are $K$ sets of high-order graph $\mathcal{B}_k^L(1 \leq k \leq K)$, each of which starts from a behavior-specified adjacency matrix $\mathbf{A}^k$.
By selecting different behavior-specific graph $\mathbf{A}^k$ at each step, we can construct a high-order graph set that contains multiple $l$-th order interaction graph with different semantics.
Specifically, the number of all possible $l$-th order graph can be calculated as:
\begin{equation}
\setlength{\abovedisplayskip}{0.5pt}
\setlength{\belowdisplayskip}{0.5pt}
card(\mathcal{B}_k^l) = pow(K, l-1)
\end{equation}
where $card(\cdot)$ is a measure of the number of elements in a set, $pow(K, l-1)$ is the function that calculates the $l-1$ power of a given number $K$, $K$ is the number of behavior types.
However, as the number of all possible $l$-th order graph is a exponential function of $l-1$, it's impractical to use such an extensive space for $l$-order interaction graph generation.

\subsubsection{Graph Compression.}
In order to find an applicable solution with limited time and space complexity, we employ a graph compression layer to construct the high-order graph sets iteratively with the node-wise multi-head attention mechanism.
The graph compression layer for target node $v$ (node $v$ could be a user node $u$ or an item node $i$) can be formulated as:
\begin{equation}
\label{equ:8}
\mathcal{B}_{v,k}^l = \mathcal{B}_{v,k}^{l-1} \otimes \{ \mathbf{\alpha}_{v,k}^{l,1} \cdot [\mathbf{A}^1, \cdots,\mathbf{A}^K], \cdots, \mathbf{\alpha}_{v,k}^{l,H}  \cdot [\mathbf{A}^1, \cdots,\mathbf{A}^K]\}
\end{equation}
\normalsize
where $\mathcal{B}_{v,k}^1 = \{\mathbf{A}^k\}$. $(\cdot)$ is the vector multiplication operation, $H$ is the number of heads and $\mathbf{\alpha}_{v,k}^{l,h} \in \mathbb{R}^{1 \times K}$ is the learned attention vector for node $v$ in the $l$-th order and the $h$-th head.
By using the node-wise multi-head attention mechanism, the number of generated $l$-th order graph is reduced from $pow(K, l-1)$ to $pow(H, l-1)$.
Since $H$ is usually much smaller than $K$ and $l-1$ is usually a very small value, so the scale of $pow(H, l-1)$ is acceptable.
The attention mechanism not only serves as a tool to reduce complexity, but is also used for finding the most useful behaviors for high-order graph generation.
To adaptively select the most relevant behavior of users and items for representation learning, we use the node-wise attention mechanism to obtain the soft weights for different behaviors, which can be defined as:
\begin{equation}
\setlength{\abovedisplayskip}{0.5pt}
\setlength{\belowdisplayskip}{0.5pt}
\mathbf{\alpha}_{u,k}^{l,h} = \sigma(\mathbf{W}_k^{l,h}\mathbf{x}_u + \mathbf{b}_k^{l,h}),
\mathbf{\alpha}_{i,k}^{l,h} = \sigma(\mathbf{W}_k^{l,h}\mathbf{y}_i + \mathbf{b}_k^{l,h})
\end{equation} 
where $\sigma(\cdot)$ is the activation function set as LeakyReLU here for better performance.
$\mathbf{W}_k^{l,h} \in \mathbb{R}^{K \times d}$ and $\mathbf{b}^{l,h} \in \mathbb{R}^{K \times 1}$ are feature transformation matrix and bias matrix, respectively.
Noticed that we also use a behavior- and layer-wise (i.e., we use different transformation matrices for different layers and behaviors) attention mechanism here, we empirically verify its effectiveness in Section \ref{attention_module}.
In this way, we can generate the personalized high-order graph sets for both users and items, which are used for later information propagation and integration.

\subsubsection{Graph Convolution}
After generating the graph set by graph interaction and graph compression layers, we enrich the representation of users and items with graph convolution.
The neighbor information propagation in each graph can be formulated as:
\begin{equation}
\setlength{\abovedisplayskip}{0.5pt}
\setlength{\belowdisplayskip}{0.5pt}
\mathbf{x}_{N_u,k}^{l,s} = Agg(\mathbf{x}_u, \mathbf{B}_{u,k}^{l,s}),
\mathbf{y}_{N_i,k}^{l,t} = Agg(\mathbf{y}_i, \mathbf{B}_{i,k}^{l,t})
\end{equation} 
where $\mathbf{B}_{u,k}^{l,s}$ and $\mathbf{B}_{i,k}^{l,t}$ denote the adjacent matrices of the $s$-th and $t$-th graph in graph set $\mathcal{B}_{u,k}^{l}$ and $\mathcal{B}_{i,k}^{l}$,  $N_u$ and $N_i$ denote the neighbors of $u$ and $i$, and $\mathbf{x}_{N_u,k}^{l,s}$ and $\mathbf{y}_{N_i,k}^{l,t}$ denote the outputs by aggregating neighbor information from $s$-th and $t$-th graph.
$Agg(\cdot)$ is a arbitrary graph convolution operator that can be used for information aggregation.
We implement $Agg(\cdot)$ with the following four state-of-the-art GCN models: \textsl{GCN Aggregator} \cite{gcn}, \textsl{NGCF Aggregator} \cite{ngcf}, \textsl{LR-GCCF Aggregator} \cite{lr-gccf}, and \textsl{LightGCN Aggregator} \cite{lightgcn}.
Notice that the matrix multiplications lead to a very dense high-order graph which is computationally unacceptable.
Therefore, we use the matrix associative property to accelerate the aggregation process for computational efficiency.
For example, $(\mathbf{A}^k \otimes  \mathbf{A}^k \otimes \mathbf{A}^k) \times \mathbf{x}_u$ can be accelerated by $\mathbf{A}^k \times  (\mathbf{A}^k \times (\mathbf{A}^k \times \mathbf{x}_u))$, where ($\times)$ is the multiplication between sparse matrix and vector.
As ($\times)$ combines a sparse matrix and a vector into a single vector, computation complexities of subsequent multiplications can be effectively reduced.
After the neighbor information propagation process, we have $pow(H, l-1)$ neighbor representations for each layer and for each node $u$ and $i$.
For simplicity, we apply the sum operation over these representations to get the final user and item representations:
\begin{equation}
\setlength{\abovedisplayskip}{0.5pt}
\setlength{\belowdisplayskip}{0.5pt}
\mathbf{x}_{N_u,k}^{l} = \sum_{s=1}^{pow(H, l-1)}\mathbf{x}_{N_u,k}^{l,s},
\mathbf{y}_{N_i,k}^{l} = \sum_{t=1}^{pow(H, l-1)}\mathbf{y}_{N_i,k}^{l,t}.
\end{equation}
To better explore high-order neighbor information and alleviate the
over-smoothing issue, we introduce the residual operation to our graph convolution layer for final node information updating, which is defined as:
\begin{equation}
\setlength{\abovedisplayskip}{0.5pt}
\setlength{\belowdisplayskip}{0.5pt}
\mathbf{x}_{u,k}^{l} = \mathbf{x}_{N_u,k}^{l} + \mathbf{x}_{u,k}^{l-1},
\mathbf{y}_{i,k}^{l} = \mathbf{y}_{N_i,k}^{l} + \mathbf{y}_{i,k}^{l-1}
\end{equation}
As the outputs of different layers reflects the relations of different orders, we finally aggregate these outputs into a single vector with the sum operation as follows:
\begin{equation}
\setlength{\abovedisplayskip}{0.5pt}
\setlength{\belowdisplayskip}{0.5pt}
\mathbf{x}_{u,k}^{*} = \sum_{l=0}^{L}{\mathbf{x}_{u,k}^{l}},
\mathbf{y}_{i,k}^{*} = \sum_{l=0}^{L}{\mathbf{y}_{i,k}^{l}}
\end{equation}
where $\mathbf{x}_{u,k}^{0} = \mathbf{x}_u$ and $\mathbf{y}_{i,k}^{0} = \mathbf{y}_i$ are the initial embeddings for user $u$ and item $i$. 
It is noticed that the central nodes aggregate neighbor information of different layers directly, which has been verified to be useful to address the heterogeneity of the user-item interaction graph \cite{nia-gcn}, compared with recursively updating the node embedding at $l$-th layer with the
output from $l-1$-th layer.


\subsection{Multi-Expert with Separate Input}\label{mesi}
With the design of CIGCN, we have obtained $K$ representations $\mathbf{x}_{u,k}^{*}$ and
$\mathbf{y}_{i,k}^{*}(1\leq k \leq K)$ for each user $u$ and each item $i$, as shown in Figure \ref{fig:framework}.
Each representation describes the personalized preferences of user $u$ or item $i$ to relations start from behavior $k$.
To alleviate the potential gradient conflict when treating multi-behavior data "\textsl{as labels}", we propose a Multi-Expert with Separate Input (MESI) network with a novel \textbf{\textsl{separate input}} design in this section.

Existing multi-behavior methods like NMTR \cite{nmtr}, GHCF \cite{ghcf} and MB-GMN \cite{mbgmn} optimize multiple tasks with the same static weights for all samples, which are limited by the sample differences, as analyzed in Section \ref{Multi-behavior_Recommendation}.
To address this problem, we use a hierarchical neural architecture which is similar to the MMOE \cite{mmoe} and PLE \cite{PLE} for MTL.
Specifically, we use experts to replace the shared bottom layer used in NMTR, GHCF and MB-GMN to learn behavior-aware information.
In this paper, each expert is defind as the combination of $\mathbf{x}_{u,k}^{*}$ and $\mathbf{y}_{i,k}^{*}$, which can be formulated as:
\begin{equation}
\setlength{\abovedisplayskip}{0.5pt}
\setlength{\belowdisplayskip}{0.5pt}
\mathbf{f}_{u,i}^{k} = \mathbf{x}_{u,k}^{*} \circ \mathbf{y}_{i,k}^{*}
\end{equation}
where ($\circ$) is the hadamard product operation, $\mathbf{x}_{u,k}^{*}$ and $\mathbf{y}_{i,k}^{*}$ are the behavior $k$ related inputs.
As the \textbf{\textsl{separate input}} are utilized here for the generation of experts, we can obtain $K$ experts in total.

As different experts may contain different preferences of users or properties of items, it's necessary to combine these experts for the final prediction of each task.
We then use the \textbf{\textsl{separate input}} to produce task-aware gate for each task to automatically select a subset of experts which are useful for the prediction of this task.
The gate for task $k$ can be defined as:
\begin{equation}
\setlength{\abovedisplayskip}{0.5pt}
\setlength{\belowdisplayskip}{0.5pt}
\mathbf{g}_{u,i}^{k} = Softmax(\mathbf{W}_g(\mathbf{x}_{u,k}^{*}||\mathbf{y}_{i,k}^{*}) + \mathbf{b}_g)
\end{equation}
where $(||)$ is the vector concatenation operation, $\mathbf{W}_g \in \mathbb{R}^{K \times 2d}$ and $\mathbf{b}_g \in \mathbb{R}^{K \times 1}$ are feature transformation matrix and bias matrix, and $\mathbf{g}_{u,i}^{k} \in \mathbb{R}^{K \times 1}$ is the attention vector which are used as selector to calculate the weighted sum of all experts.
The final prediction score for task $k$ is calculated as:
\begin{equation}
\setlength{\abovedisplayskip}{0.5pt}
\setlength{\belowdisplayskip}{0.5pt}
\hat{o}_{u,i}^{k} = h^{k}(\sum_{j=1}^{K} {\mathbf{g}_{u,i}^{k}(j) \cdot \mathbf{f}_{u,i}^{j}})
\end{equation}
where $\mathbf{g}_{u,i}^{k}(j)$ denotes the $j$-th element of vector $\mathbf{g}_{u,i}^{k}$, $h^{k}(\cdot)$ is the tower function. 
Following \cite{moe}, we use average operation as the tower function here for simplicity. 

\subsection{Joint Optimization for MTL}
Since we have obtained the prediction value $\hat{o}_{u,i}^{k}$ for each type of behavior $k$, we use the  \textit{Bayesian Personalized Ranking} (BPR) \cite{bpr} loss for multi-task learning, which can be formulated as:
 
\begin{equation}
\setlength{\abovedisplayskip}{0.5pt}
\setlength{\belowdisplayskip}{0.5pt}
\mathcal{L} = -\sum_{k=1}^{K}\sum_{(u,s,t)\in \mathcal{O}_k} \textup{ln} \sigma(\hat{o}_{u,s}^{k} - \hat{o}_{u,t}^{k}) + \lambda ||\Theta||^2_2
\end{equation}
where $\mathcal{O}_k = \left\{(u,s,t)|(u,s)\in \mathcal{O}_k^{+}, (u,t) \in \mathcal{O}_k^{-} \right\}$ denotes the training dataset. $\mathcal{O}_k^+$ indicates observed positive user-item interactions under behavior $k$ and $\mathcal{O}_k^-$ indicates unobserved user-item interactions under behavior $k$. $\Theta$ represents set of all model parameters, $\sigma$ is the Sigmoid function and $\lambda$ is the $L_2$ regularization coefficient for $\Theta$.

\begin{table}[t]
\setlength{\abovecaptionskip}{-0cm}
\setlength{\belowcaptionskip}{-0.1cm}
 \caption{\small{Dataset statistics.}}
 \centering
 	\setlength{\tabcolsep}{1mm}
 \begin{tabular}{@{} c|c|c|c|c @{}}
 \hline
 \textbf{Dataset} 	  & \textbf{User}   & \textbf{Item} & \textbf{Interaction} & \textbf{Behaviors}  \\
 \hline
 Beibei  & 21,716 & 7,977 & 3,338,068 & View,Cart,Buy \\
 Taobao & 147,894 & 99,037 & 7,658,926 & View,Favorite,Cart,Buy \\
 IJCAI & 423,423 & 874,328 & 36,203,512 & View,Favorite,Cart,Buy \\ 
 \hline
\end{tabular}
\label{tab:dataset}
\vspace{-5mm}
\end{table}





%% file: sections/experiments.tex
\section{EXPERIMENTS}\label{experiment}
\subsection{Experiment Setup}\label{ExperimentSetup}

\subsubsection{Datasets}

	


To reduce biases, we adopt the same public datasets (i.e., \textbf{Beibei}, \textbf{Taobao}, and \textbf{IJCAI})\footnote{https://github.com/akaxlh/MB-GMN} and pre-processings as in MB-GMN \cite{mbgmn}, and the statistics are shown in Table \ref{tab:dataset}. 


 
\subsubsection{Compared Baseline}

For a comprehensive comparison, we compare CIGF against four types of representative baselines: i) NNs-based single-behavior models, i.e., DMF \cite{deepmf} and AutoRec \cite{autorec}; ii) NNs-based multi-behavior models, i.e., NMTR \cite{nmtr}, DIPN \cite{dipn}, and MATN \cite{matn}; iii) GNNs-based single-behavior models, i.e., NGCF \cite{ngcf} and LightGCN \cite{lightgcn}; iv) GNNs-based multi-behavior models, i.e., NGCF$_M$ \cite{ngcf}, LightGCN$_M$ (LightGCN \cite{lightgcn} enhanced with the multi-behavioral graph), GHCF \cite{ghcf}, and MBGCN \cite{mbgcn}.
Public codes for GHCF\footnote{https://github.com/chenchongthu/GHCF; Due to the unaffordable memory usage brought by non-sampling learning, GHCF is inapplicable to the IJCAI dataset.} and LightGCN\footnote{https://github.com/kuandeng/LightGCN} are used, while the best results for other models (DMF, AutoRec, NGCF, NMTR, DIPN, MATN, MBGCN, and MB-GMN) are picked from \cite{mbgmn}.

\subsubsection{Evaluation Metrics}
The Hit Ratio (HR@$N$) and Normalized Discounted Cumulative
Gain (NDCG@$N$) are used to evaluate the performances.
By default, we set $N = 10$ in all experiments. Similar results of other metrics (i.e., $N = 1, 5, 20$) on the three datasets can also be obtained, whereas they are not presented here due to the space limitation. And the details of implementation are shown in Appendix \ref{parameters}.

\subsection{Overall Performance Comparison}\label{PerformanceComparison}
From Table \ref{tab:ctraccuracy}, we have the following observations in terms of model effectiveness (analysis of complexity is shown in Appendix \ref{complexity and efficiency}):
\begin{table}[h]
\setlength{\abovecaptionskip}{0cm}
\setlength{\belowcaptionskip}{-0.1cm}
\centering
\caption{The overall comparison. 
$\star$ indicates a statistically significant level $p$-value<0.05 comparing CIGF with the best baseline (indicated by underlined numbers).}
\setlength{\tabcolsep}{1mm}{
\small
\begin{tabular}{c|c|c|c|c|c|c}
\midrule[0.25ex]
Dataset &
\multicolumn{2}{c|}{Beibei} & 
\multicolumn{2}{c|}{Taobao} &
\multicolumn{2}{c}{IJCAI} \\ \hline 
Model & HR & NDCG &  HR & NDCG  & HR & NDCG \\\hline \hline
DMF & 0.597 & 0.336 & 0.305 & 0.189 & 0.392 & 0.250 \\
AutoRec  & 0.607 & 0.341 & 0.313 & 0.190 & 0.448 & 0.287\\\hline
NGCF & 0.611 & 0.375 & 0.302 & 0.185 & 0.461 & 0.292 \\
LightGCN & 0.643 & 0.378 & 0.373 & 0.235 & 0.443 & 0.283 \\\hline
NMTR  & 0.613 & 0.349 & 0.332 & 0.179 & 0.481 & 0.304   \\
DIPN & 0.631 & 0.394 & 0.317 & 0.178 & 0.475 & 0.296  \\ 
MATN & 0.626 & 0.385 & 0.354 & 0.209 & 0.489 & 0.309    \\
\hline
NGCF$_M$ & 0.634 & 0.372 & 0.374 & 0.221 & 0.481 & 0.307 \\
LightGCN$_M$ & 0.651 & 0.391 & 0.391 & 0.243 & 0.486 & 0.317 \\
GHCF & 0.608 & 0.378 & 0.415 & 0.241 & - & -\\
MBGCN & 0.642 & 0.376 & 0.369 & 0.222 & 0.463 & 0.277 \\
MB-GMN & \underline{0.691} & \underline{0.410} & \underline{0.491} & \underline{0.300} & \underline{0.532} & \underline{0.345} \\ \hline
CIGF & $\textbf{0.700}^{\star}$ & $\textbf{0.443}^{\star}$ & $\textbf{0.592}^{\star}$ & $\textbf{0.383}^{\star}$ & $\textbf{0.601}^{\star}$ & $\textbf{0.400}^{\star}$ \\
\hline \hline
\%Improv & 1.30\% & 8.05\% & 20.57\% & 27.67\% & 12.97\%  & 15.94\%  \\\hline
\end{tabular}}
\label{tab:ctraccuracy}
\vspace{-4mm}
\end{table}
\begin{itemize}
    \item CIGF consistently yields superior performance on all three datasets. More precisely, CIGF outperforms the strongest baselines by \textbf{1.30\%}, \textbf{20.57\%}, and \textbf{12.97\%} in terms of \textit{HR} (\textit{8.05\%}, \textit{27.67\%}, and \textit{15.94\%} in terms of \textit{NDCG}) on Beibei, Taobao, and IJCAI, respectively.
    Additionally, the performance improvements on Taobao and Ijcai datasets are much more significant than that on Beibei dataset.
    One possible reason is that the interaction information of different behaviors contained in Beibei dataset is mutually covered (as shown in Appendix \ref{Label_Correlations}, users who have bought an item must also have viewed and carted it), which reduces the significance of high-order relation modeling. 
     \item NGCF and LightGCN perform better than DMF and AutoRec on most datasets, which demonstrates the advantage of GNN in extracting high-order collaborative signals.
     By distinguishing different behaviors, NMTR, DIPN, and MATN achieve much better performances than DMF and AutoRec.
     This verifies the necessity to extract and model the relation information between different types of behaviors. 
    \item NGCF, LightGCN, NMTR, DIPN, and MATN perform worse than NGCF$_M$, LightGCN$_M$, MBGCN, and MB-GMN on most datasets, which indicates the incapability of NNs models and single-behavior GNNs models in modeling high-order multi-behavior relations.
    This justifies the necessity to simultaneously consider multi-behavior and high-order relations.

\end{itemize}



\subsection{Ablation Study of CIGF}



\subsubsection{On the effectiveness of key components}
\begin{table}[h]
\setlength{\abovecaptionskip}{0cm}
\setlength{\belowcaptionskip}{-0.1cm}
\caption{Performances of different CIGF variants.}
    \centering
\setlength{\tabcolsep}{1mm}{
\small
\begin{tabular}{c|c|c|c|c|c|c}
\midrule[0.25ex]
Dataset &
\multicolumn{2}{c|}{Beibei} & 
\multicolumn{2}{c|}{Taobao} &
\multicolumn{2}{c}{IJCAI} \\ \hline 
Model & HR & NDCG &  HR & NDCG  & HR & NDCG \\\hline \hline
Base Model & 0.649  & 0.392  & 0.444  & 0.275  & 0.457 & 0.297    \\
w/o CIGCN  & 0.660  & \underline{0.410}  & 0.460  & 0.286  & 0.495  & 0.322  \\
w/o MESI & \underline{0.662}  & 0.401  & \underline{0.528}  & \underline{0.340}   & \underline{0.573} & \underline{0.382}    \\ 
CIGF   & \textbf{0.700}  & \textbf{0.443}  & \textbf{0.592}  & \textbf{0.383}  & \textbf{0.601} & \textbf{0.400}   \\
\hline \hline
\end{tabular}}
\vspace{-3mm}
\label{tab:Impact of main parts}
\end{table}
To evaluate the effectiveness of sub-modules in our CIGF framework, we consider three model variants:
 (1) \textbf{\textsl{Base Model}:} 
    We remove CIGCN part (i.e., the behavior-specific graph are used for convolution directly) and replace the MESI network with bilinear module.
    This variant cannot model \textbf{\textsl{instance-level}} high-order relations and use \textbf{\textsl{same input}} for MTL.
     (2) \textbf{\textsl{w/o CIGCN}:} 
    The CIGCN part is removed.
    (3) \textbf{\textsl{w/o MESI}:}
    The MESI part is replaced with bilinear module.
    As shown in Table \ref{tab:Impact of main parts}, both CIGCN and MESI bring performance improvements compared with base model, and the complete CIGF framework achieves the best results.
    Therefore, we claim that both \textbf{\textsl{instance-level}} high-order multi-behavior relation and \textbf{\textsl{separate input}} are effective and complementary to each other.
    And it's necessary to treat multi-behavior data both "\textsl{as features}" and "\textsl{as labels}".


\label{attention_module}
\begin{table}[H]
\setlength{\abovecaptionskip}{-0cm}
\setlength{\belowcaptionskip}{-0.1cm}
\caption{Performances of different attention variants.}
    \centering
\setlength{\tabcolsep}{1mm}{
\small
\begin{tabular}{c|c|c|c|c|c|c}
\midrule[0.25ex]
Dataset &
\multicolumn{2}{c|}{Beibei} & 
\multicolumn{2}{c|}{Taobao} &
\multicolumn{2}{c}{IJCAI} \\ \hline 
Method & HR & NDCG &  HR & NDCG  & HR & NDCG \\\hline \hline
global-wise  & 0.694 & 0.438 & 0.565 & 0.361 & 0.572 & 0.373  \\
node-wise   & 0.698 & 0.442 & 0.561 & 0.356 & 0.587 & 0.390 \\
node-wise+layer   & 0.698 & \underline{0.442} & 0.564 & 0.361 & 0.590 & 0.392 \\
node-wise+beh   & \underline{0.698} & 0.441 & \underline{0.588} & \underline{0.379} & \underline{0.599} & \underline{0.398} \\
node-wise+beh+layer  & \textbf{0.700}  & \textbf{0.443}  & \textbf{0.592}  & \textbf{0.383}  & \textbf{0.601} & \textbf{0.400} \\
\hline \hline
\end{tabular}}
\vspace{-4mm}
\label{tab:Impact of different attention}
\end{table}
\label{relation_attention}

\subsubsection{On the impact of attention module.}

To demonstrate the effectiveness of our attention module, we consider four variants:
(1) \textbf{\textsl{global-wise}:} The attention weight is global-wise for all user/item.
(2) \textbf{\textsl{node-wise}:} The attention weight is shared by all layers and behaviors but different for each user/item.
(3) \textbf{\textsl{node-wise+layer}:} The attention weight is shared by all behaviors but different for each layer or user/item.
(4) \textbf{\textsl{node-wise+beh}:} The attention weight is shared by all layers but different for each behavior or user/item.
From the results displayed in Table \ref{tab:Impact of different attention},
global-wise attention performs the worst among all variants in most cases, which suggests the importance of learning the customized information for each node. 
Besides, all the enhanced node-wise variants perform better than pure node-wise attention mechanism, and our proposed attention mechanism achieves the best performance on all three datasets.
The results indicate the effectiveness and rationality of our proposed  behavior-wise and layer-wise node-wise attention mechanism for high-order multi-behavior relation selection.



\subsubsection{On the impact of MTL modules.}
\begin{table}[t]
\setlength{\abovecaptionskip}{0cm}
\setlength{\belowcaptionskip}{0cm}
\caption{Impact of MTL modules.}
    \centering
\setlength{\tabcolsep}{1mm}{
\small
\begin{tabular}{c|c|c|c|c|c|c}
\midrule[0.25ex]
Dataset &
\multicolumn{2}{c|}{Beibei} & 
\multicolumn{2}{c|}{Taobao} &
\multicolumn{2}{c}{IJCAI} \\ \hline 
Method & HR & NDCG &  HR & NDCG  & HR & NDCG \\\hline \hline
CIGCN-SB & 0.605  & 0.341  & 0.389  & 0.231  & 0.485 & 0.302   \\
CIGCN-Bilinear & 0.662 & \underline{0.401} & 0.528  & \underline{0.340}  & \underline{0.573} & \underline{0.382} \\
CIGCN-MMOE  & \underline{0.663}  & 0.391  & \underline{0.541}  & 0.338  & 0.546 & 0.339 \\
CIGCN-PLE   & 0.653  & 0.381  & 0.521  & 0.320  & 0.526 & 0.325   \\
CIGF  & \textbf{0.700} & \textbf{0.443}  & \textbf{0.592}  & \textbf{0.383} & \textbf{0.601} & \textbf{0.400}  \\
\hline \hline
\end{tabular}}
\label{tab:Impact of multi-beh learning}
\vspace{-3mm}
\end{table}

To further demonstrate the superiority of our proposed MESI for MTL, we replace it with four state-of-the-art MTL modules, namely, Shared Bottom \cite{sharebottom}, Bilinear \cite{ghcf}, MMOE \cite{mmoe}, and PLE \cite{PLE}, and apply them on the top of CIGCN for multi-behavior recommendation.
Notice that there are $K$ representations used as $\textbf{\textsl{separate input}}$ generated from CIGCN.
To make it applicable for these four modules which use $\textbf{\textsl{same input}}$, we average the $K$ representations to get one unified input. 
Resulted variants are named as CIGCN-SB, CIGCN-Bilinear, CIGCN-MMOE, CIGCN-PLE, and CIGF respectively.
The results are summarized in Table \ref{tab:Impact of multi-beh learning}.
As we can see, CIGCN-SB performs the worst among all MTL models on all datasets.
CIGCN-Bilinear replaces the prediction head of neural network in CIGCN-SB with lightweight matrix transformation and performs better.
Possible reason is that the lightweight operation can reduce the risk of overfitting.
Besides, both CIGCN-MMOE and CIGCN-PLE have employed the gate network with adaptive attention weights for information fusing, thus outperform the static and same-weighted CIGCN-SB.
Finally, our MESI consistently performs the best on all datasets.
This verifies the effectiveness of $\textbf{\textsl{separate input}}$ for MTL.


\subsubsection{On the impact of GCN aggregators.}\label{Imapct_of_aggregatoe}

\begin{figure}[t]
	\setlength{\belowcaptionskip}{-0.3cm}
	\setlength{\abovecaptionskip}{0cm}
	\subfigure{
        \begin{minipage}[t]{0.48\linewidth}
        \centering
		\label{fig:agg_hr} 
		\includegraphics[width=1.6in]{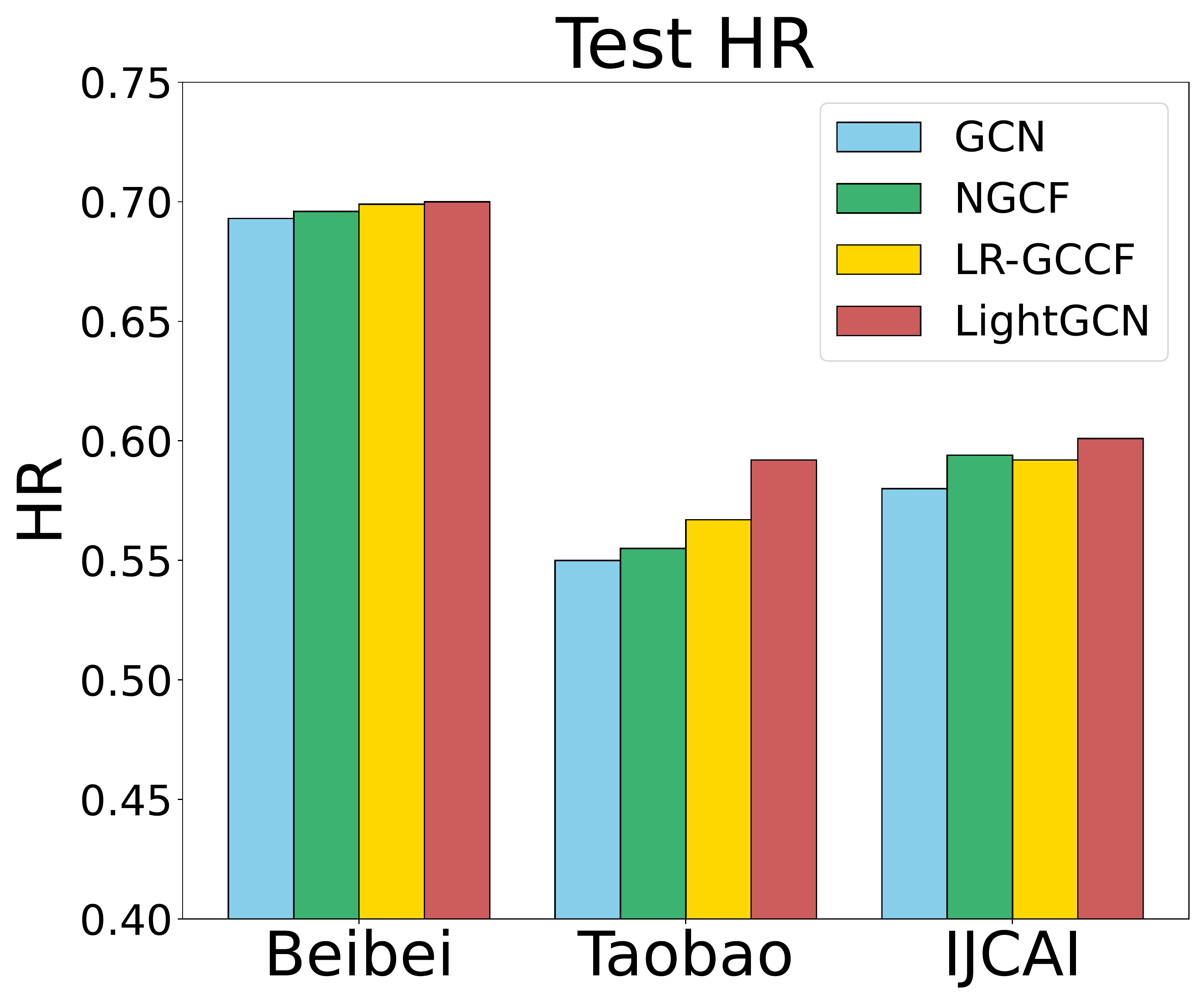}
        \end{minipage}}
	\subfigure{
        \begin{minipage}[t]{0.48\linewidth}
        \centering
		\label{fig:agg_ndcg} 
		\includegraphics[width=1.6in]{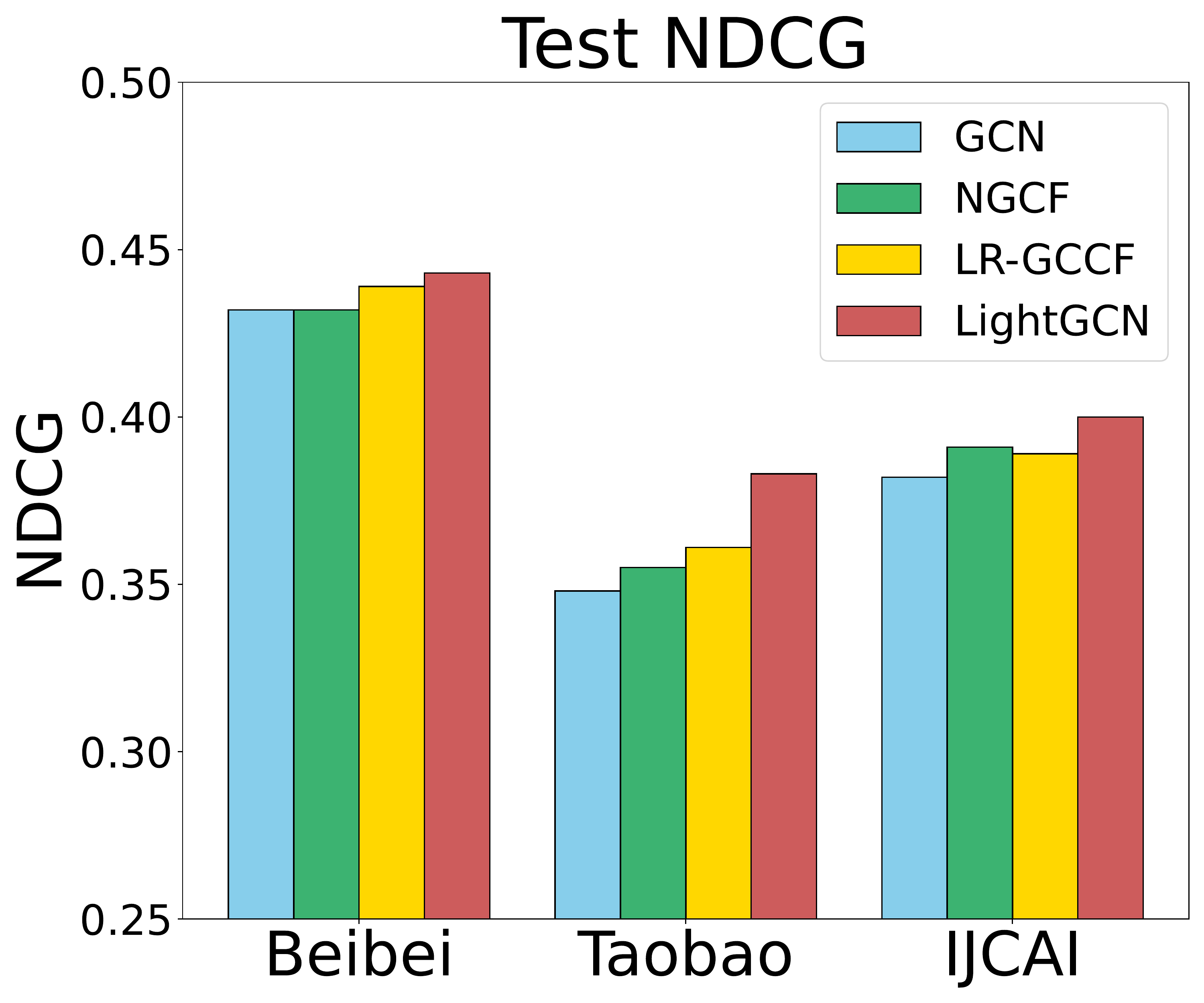}
        \end{minipage}}
	\caption{Impact of GCN aggregators.}
	\label{fig:figureaggregators}
\vspace{-3mm}
\end{figure}
To explore the impact of different GCN aggregators, we compare the variants of our proposed model with different GCN aggregators, including \textsl{GCN Aggregator} \cite{gcn}, \textsl{NGCF Aggregator} \cite{ngcf}, \textsl{LR-GCCF Aggregator} \cite{lr-gccf}, and \textsl{LightGCN Aggregator} \cite{lightgcn}.
The experimental results are illustrated in Figure \ref{fig:figureaggregators}.
We can see that NGCF Aggregator performs better than GCN aggregators on all datasets.
A possible reason is that additional feature interactions introduced by NGCF Aggregator provides more information.
We also find that LR-GCCF Aggregator performs slightly better than NGCF Aggregator on almost all datasets.
The reason is that removing the transformation matrix can alleviate overfitting.
Moreover, LightGCN Aggregator obtains the best performance on all datasets by simultaneously removing transformation matrix and activation function.

\subsection{In-depth Analysis of Model Design}
In this part, we conduct experiments to make in-depth analysis about \textbf{\textsl{instance-level}} high-order relation modeling when treating multi-behavior data "\textsl{as features}" and potenial gradient conflict when treating multi-behavior data "\textsl{as labels}".

\begin{figure}[t]
    \setlength{\belowcaptionskip}{-0.3cm}
    \setlength{\abovecaptionskip}{0cm}
    \subfigure{
        \begin{minipage}[t]{0.48\linewidth}
        \centering
        \label{fig:layer_beibei} 
        \includegraphics[width=1.6in]{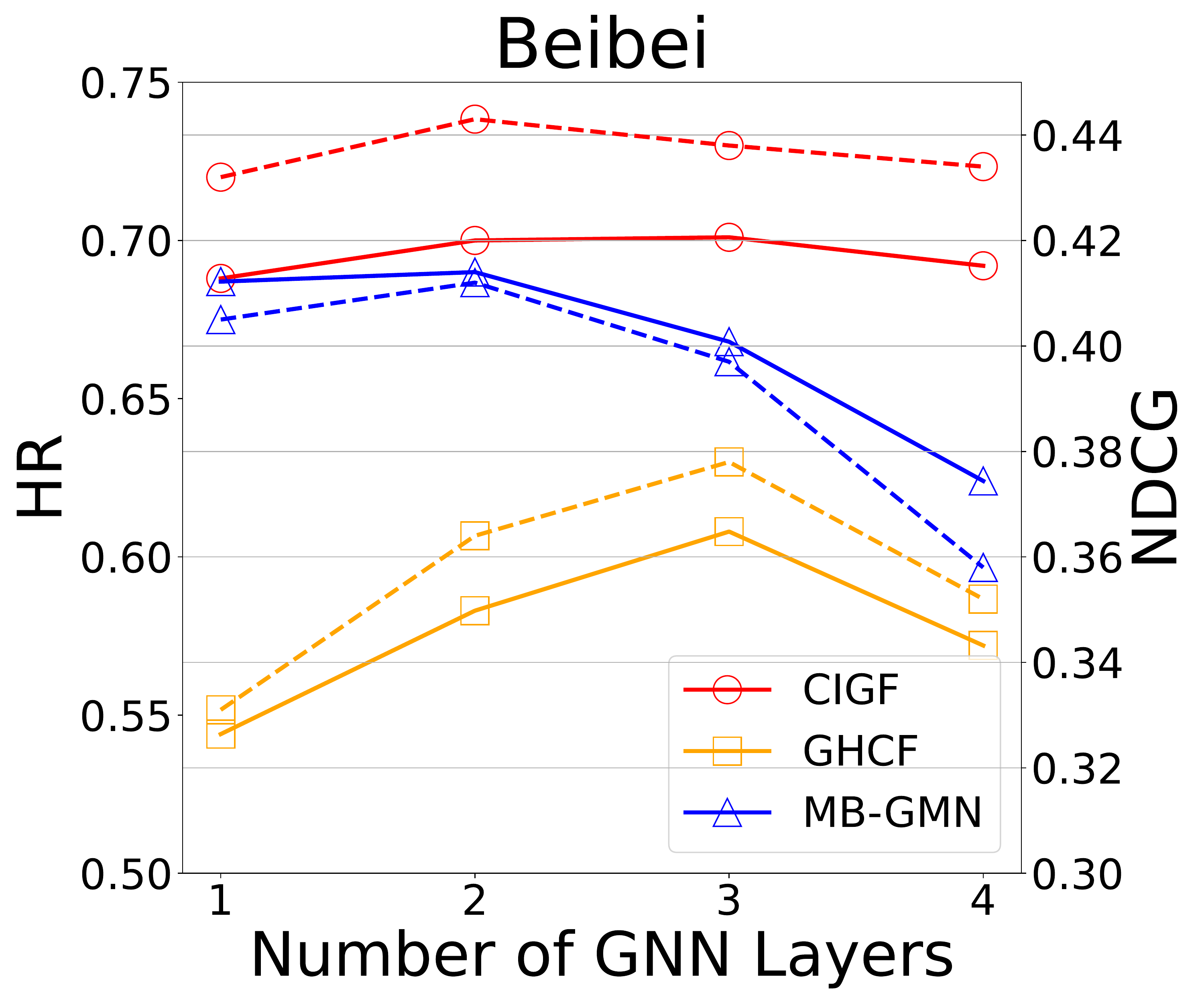}
        \end{minipage}}
    \subfigure{
        \begin{minipage}[t]{0.48\linewidth}
        \centering
        \label{fig:layer_taobao} 
        \includegraphics[width=1.6in]{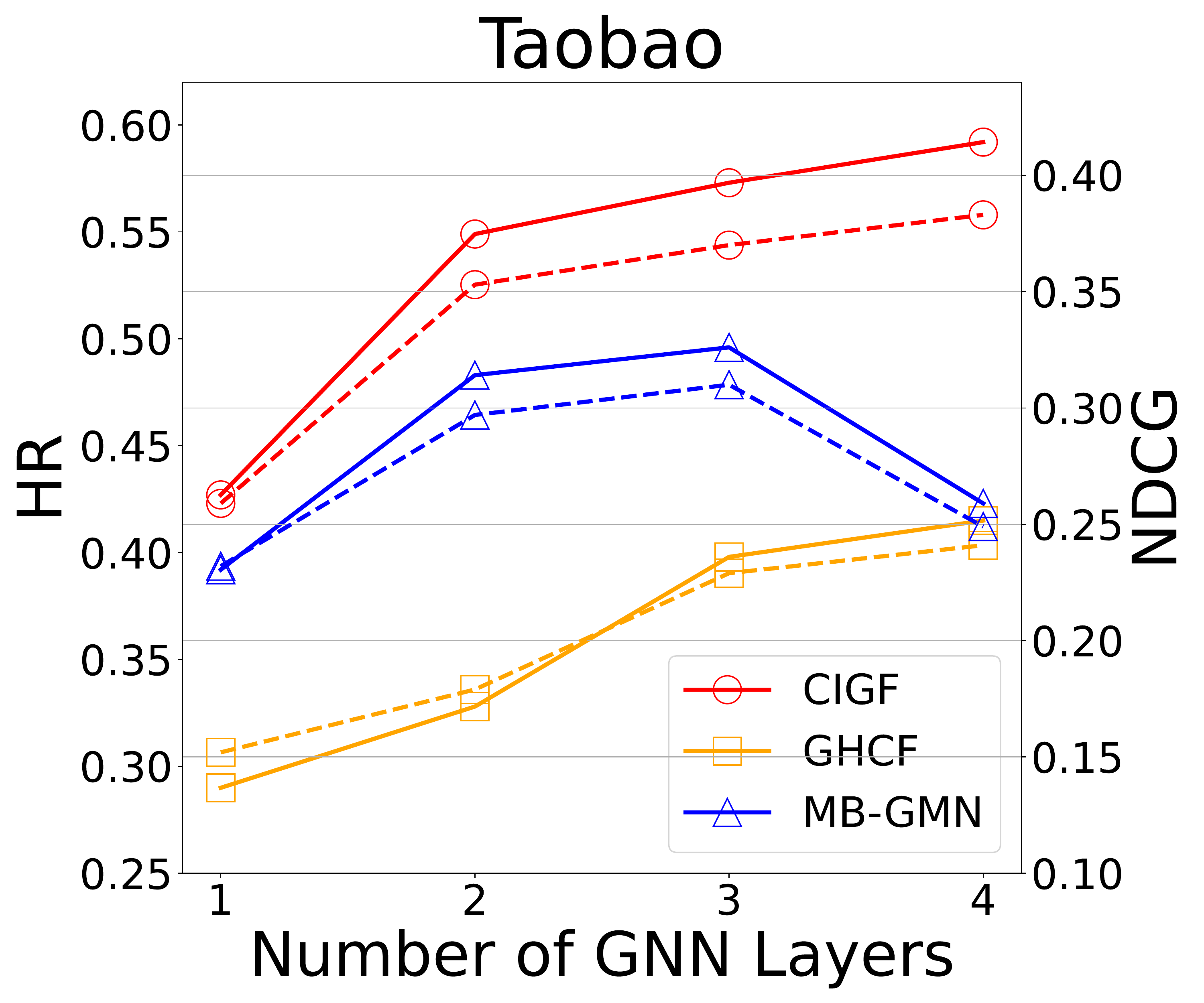}
        \end{minipage}}
    \Description{Effect of the layer number. The solid line and the dotted line represent HR and NDCG, respectively.}
    \caption{Effect of the layer number. The solid line and the dotted line represent HR and NDCG, respectively.}
    \label{fig:layers}
\end{figure}

\subsubsection{Instance-level high-order relation modeling.}

We vary the depth of CIGF to investigate whether our model can benefit from \textbf{\textsl{instance-level}} high-order relations.
And we compare the results with GHCF and MB-GMN which model \textbf{\textsl{behavior-level}} high-order relations. 
Due to lack of space, we only show the results on Beibei and Taobao datasets in Figure \ref{fig:layers}, the result of another dataset is consistent.
We can see that CIGF consistently outperforms the other methods when the layer number increases.
Besides, we can also find CIGF keeps stable on Beibei and increases continuously on Taobao, while MBGCN degrades rapidly on both datasets and GHCF degrades rapidly on Beibei when increasing the layer number. 
This observation verifies the effectiveness of our proposed method for \textbf{\textsl{instance-level}} high-order relation modeling.



\begin{table}[t]
\setlength{\abovecaptionskip}{-0cm}
\setlength{\belowcaptionskip}{-0.1cm}
\caption{The selected top-3 and bottom-3 relations.}
    \centering
\setlength{\tabcolsep}{1mm}{
\small
\begin{tabular}{c|c|c|c|c}
\midrule[0.25ex]
Dataset & Relation &
Second order &
Third order & 
Fourth order \\\hline \hline
\multicolumn{1}{c|}{\multirow{2}{*}{Beibei}} & top-3 & CC, CV, PP & - & -      \\
\multicolumn{1}{c|}{} & bottom-3 & CP, PV, VP & - & -  \\\hline
\multicolumn{1}{c|}{\multirow{2}{*}{Taobao}} & top-3 & FP, CP, PC & FPP, CPP, PCP & PFFP, CFFP, FPPV     \\
\multicolumn{1}{c|}{} & bottom-3 & PP, CC, FF & PPP, PPV, PPF & PPPF, PPPC, PPPP   \\\hline
\multicolumn{1}{c|}{\multirow{2}{*}{IJCAI}} & top-3 & CV, FV, PV & PVV, CVV, FFF & PVPV, CPPV, CVPV     \\
\multicolumn{1}{c|}{} & bottom-3 & FC, CC, PC & FCF, CCF, FCP & FCFC, CCFC, PCFC  \\
\hline \hline
\end{tabular}}
\label{tab:vis attention}
\vspace{-4mm}
\end{table}

\textbf{\textsl{Instance-level}} high-order relations bring benefits for final recommendation.
Besides, it can also reveal the underlying reasons that motivate users’ preferences on items.
Towards this end, we select and show the top-3 and bottom-3 relations among all possible relations for each order according to the average attention weights of all users in Table \ref{tab:vis attention}.
Notice that there are only second order relations on Beibei as our model achieves best results with two layers on this dataset.
As we can see, the third order relation $user \xrightarrow{favor} item \xrightarrow{be \; purchased \; by} user \xrightarrow{ purchase} item$ has the highest average attention weights on Taobao.
Possible explanation is that users tend to purchase items bought by similar users.
Besides, the second order relation $user \xrightarrow{favor} item \xrightarrow{be \; carted \; by} user$ has the lowest average attention weights on IJCAI.
The rationality can be verified in Appendix \ref{Label_Correlations} that the probability that users only have favor and cart behaviors with the same items is zero on IJCAI.



\subsubsection{Gradient conflict analysis.}

\begin{figure}[t]
	\setlength{\belowcaptionskip}{-0.4cm}
	\setlength{\abovecaptionskip}{-0.1cm}
	\subfigure{
        \begin{minipage}[t]{0.47\linewidth}
        \centering
		\label{fig:case_beibei} 
		\includegraphics[width=1.5in]{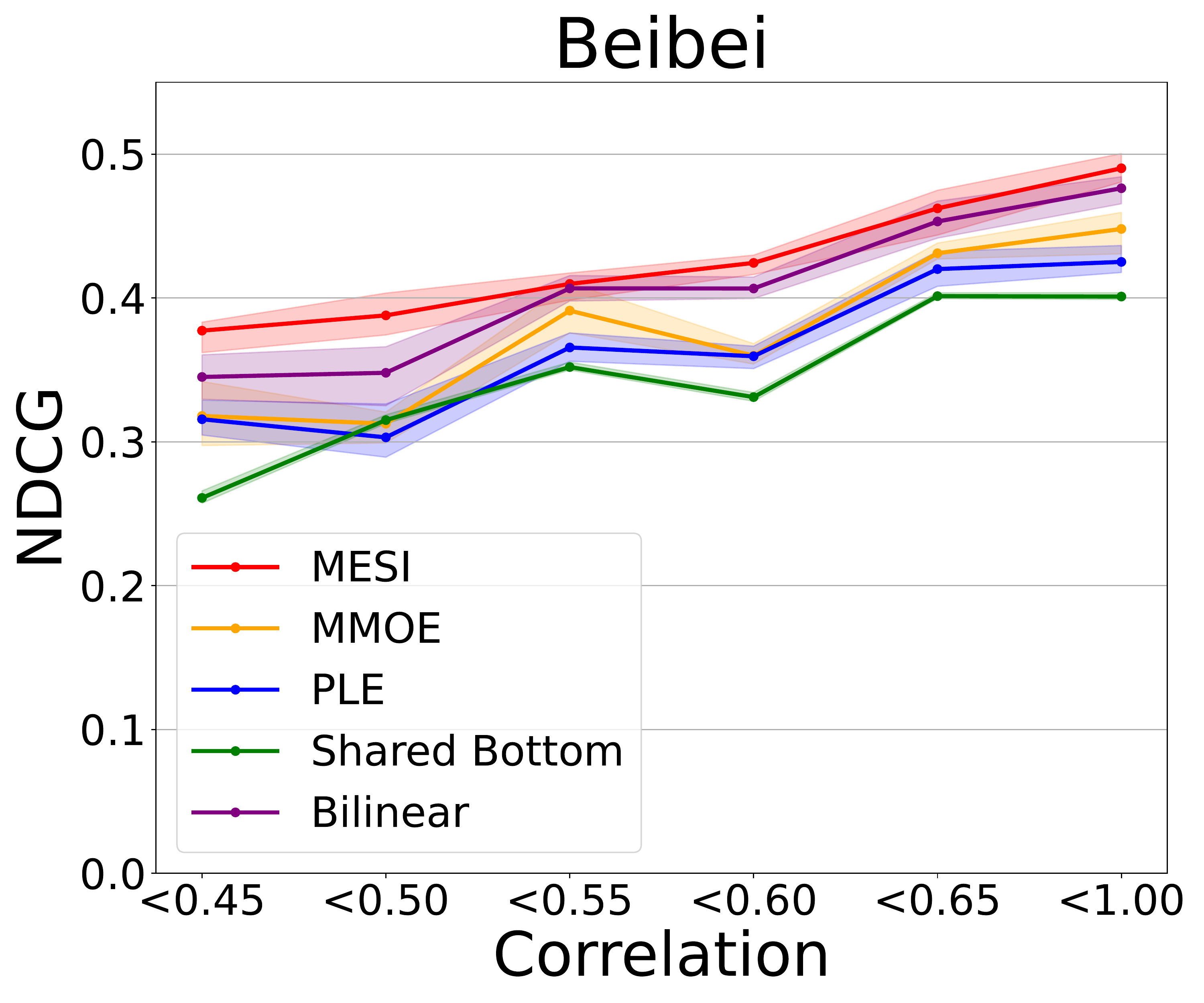}
        \end{minipage}}
	\subfigure{
        \begin{minipage}[t]{0.47\linewidth}
        \centering
		\label{fig:case_taobao} 
		\includegraphics[width=1.5in]{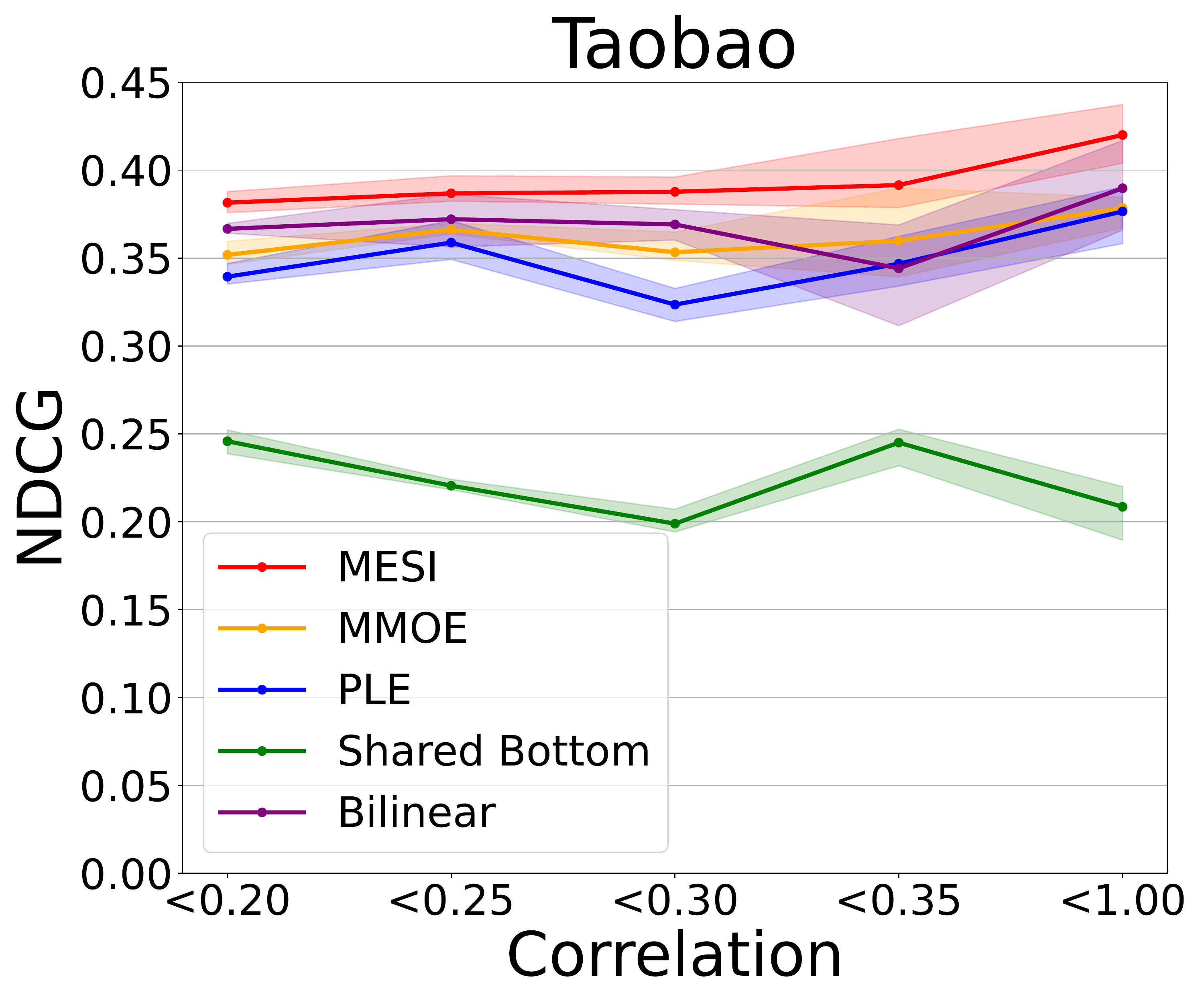}
        \end{minipage}}
    \Description{Average performances for user groups with different behavior correlations.}
	\caption{Average performances for user groups with different behavior correlations.}
	\label{fig:case_study}
\end{figure}

To verify that our model can alleviate potential gradient conflict, we perform experiments on user groups with different behavior relevance levels.
In particular, we divide the test set into six user groups according to the average Pearson correlation \cite{berthold2016clustering} among all behaviors.
The calculation of average Pearson correlation can be referred to Appendix \ref{Calculation_of_Pearson_correlation}. 
For fair comparison, we select a subset from each user group to keep the interaction number for each user fixed, thus preventing the potential impact of node degree to results \cite{ngcf}.
Figure \ref{fig:case_study} presents the results.
We omit the results on the IJCAI dataset due to space limitation, which have consistent trends.
For more rigorous results, we run each experiment 5 times and draw the mean and fluctuation range on the figure.
We find that MESI consistently outperforms all baselines among all user groups, which further demonstrates the superiority of MESI for MTL.
Besides, with the increase of behavior correlations, MESI gets better performances, while the performances of other baselines fluctuate or even decrease.
A possible reason is the negative transfer caused by potential gradient conflict when knowledge is transferred across different tasks. 
\begin{figure}[t]
    \setlength{\belowcaptionskip}{-0.3cm}
    \setlength{\abovecaptionskip}{-0.1cm}
    \subfigure{
        \begin{minipage}[t]{0.3\linewidth}
        \centering
        \label{fig:sesg} 
        \includegraphics[width=1in]{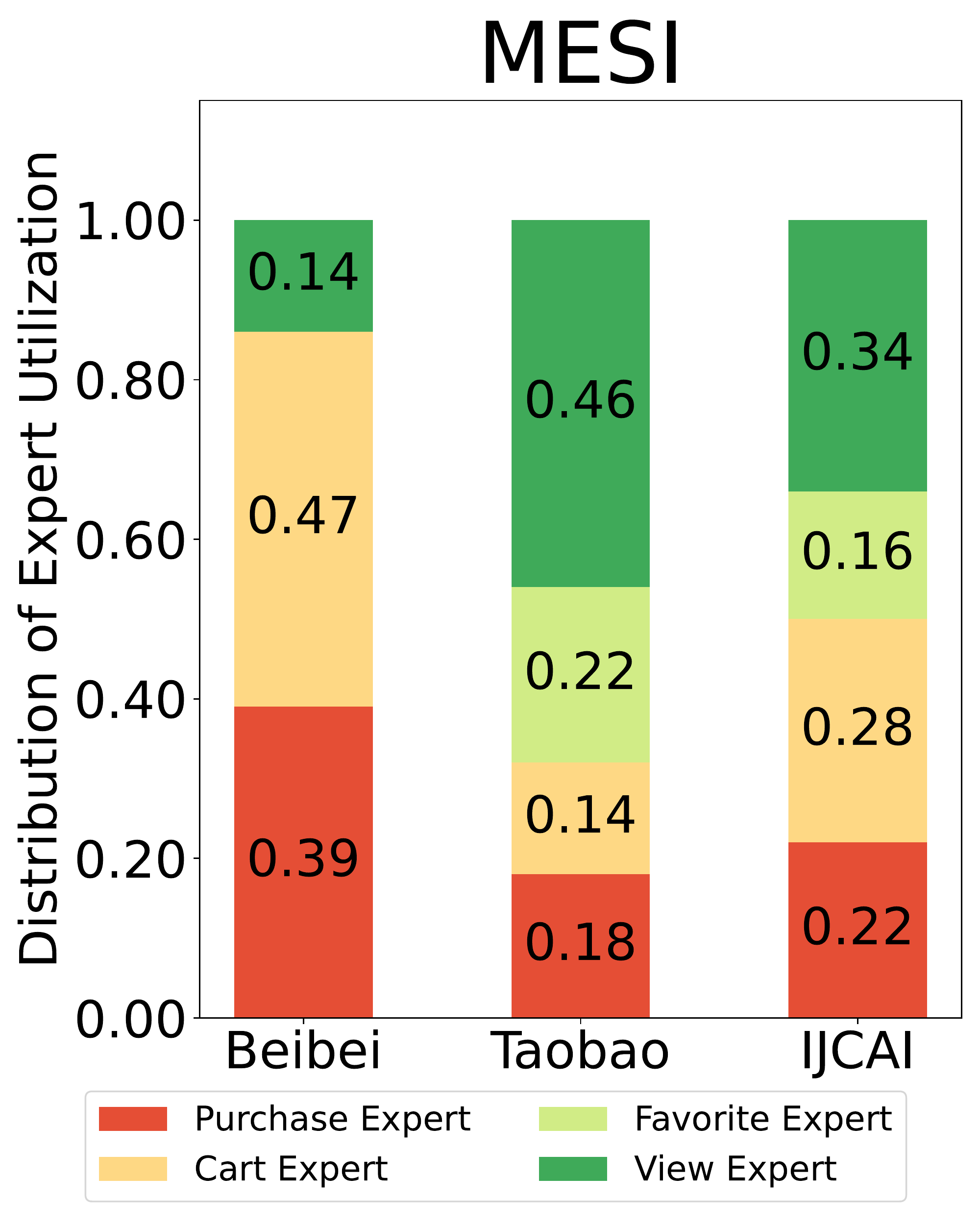}
        \end{minipage}}
    \subfigure{
        \begin{minipage}[t]{0.3\linewidth}
        \centering
        \label{fig:mmoe} 
        \includegraphics[width=1in]{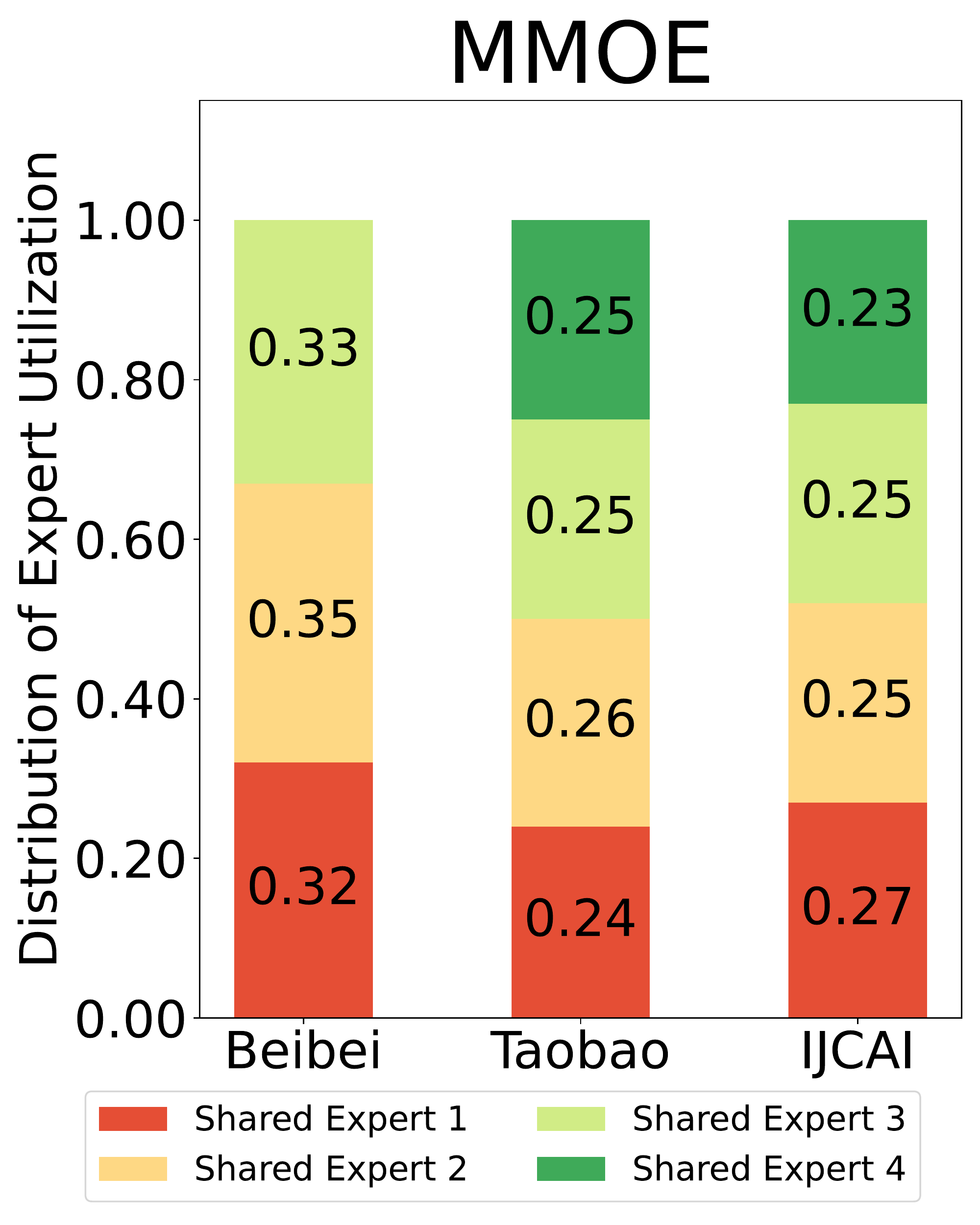}
        \end{minipage}}
    \subfigure{
        \begin{minipage}[t]{0.3\linewidth}
        \centering
        \label{fig:ple} 
        \includegraphics[width=1in]{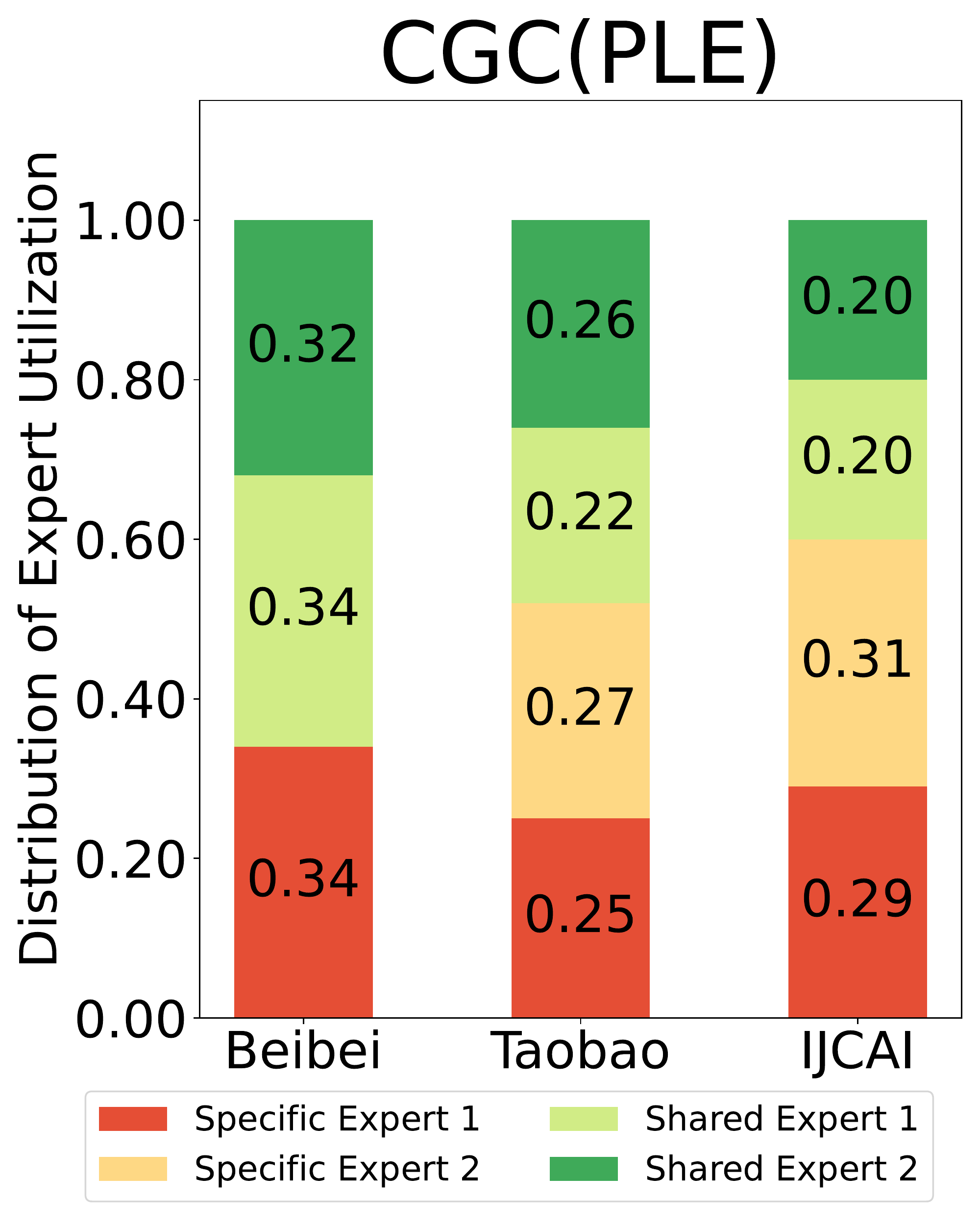}
        \end{minipage}}
    \Description{Expert utilization in gate-based models}
    \caption{Expert utilization in gate-based models}
    \label{fig:gate_distribution}
\end{figure}
To understand the reason why our proposed MESI can alleviate potential gradient conflict, we conduct experiments to compare the experts utilization among our MESI and other gate-based models (MMOE and PLE).
Following \cite{PLE}, we visualize the average weight distribution of experts used by the target behavior prediction in Figure \ref{fig:gate_distribution}.
Notice that we omit gates used for other behaviors as our goal is to predict the interaction probability of target behavior. 
Besides, for the sake of comparison, we fix the number of experts as 3 on Beibei dataset and 4 on Taobao and IJCAI datasets for both MMOE and PLE. 
It is shown that our MESI achieves better differentiation between different experts while MMOE and PLE have a nearly uniform distribution for all experts.
Thus our MESI can selectively leverage information of different behaviors to update the gradient to avoid potential conflict.

%% file: sections/conclusion.tex
\section{Conclusions}
In this paper, we propose the CIGF framework for multi-behavior recommendations.
To explicitly model \textbf{\textsl{instance-level}} high-order relations, we introduce the CIGCN module, which leverages matrix multiplication as the interaction operator to generate high-order interaction graphs, and perform graph convolution on these graphs to explore relation integration.
To alleviate potential gradient conflict, we propose the MESI network, which uses behavior-specific \textbf{\textsl{separate inputs}} explicitly. By doing so, the risk of negative transfer is reduced.
We conduct comprehensive experiments on three real-world datasets and show that the proposed CIGF outperforms all the state-of-the-art methods on all three datasets.
Further analysis shows that CIGF can fully capture high-order relationships and effectively alleviate negative transfer.


%% file: appendix.tex
\newpage

\appendix
\section{Appendix}
\subsection{Parameter Settings}
\label{parameters}
Our proposed CIGF is implemented in TensorFlow \cite{TensorFlow}. 
For a fair comparison, we set the embedding size of both users and items to 16 for all models, and initialize the model parameters with Xavier method \cite{xavier}. We adopt Adam \cite{kingma2014adam} to optimize the models and set the learning rate of 0.001 and batch size of 256, respectively.
Moreover, the number of GCN layers for graph models is searched from \{1,2,3,4,5\}.
We only use one head in the graph compression layer for simplicity as it has already achieved enough performance improvements. 
Other parameter settings are kept consistent with MB-GMN \cite{mbgmn}. 
All experiments are run for 5 times and average results are reported.




\subsection{The Coupled Gradient Issue in MTL}\label{proof_of_coupled_gradient}


For the sake of simplicity, we assume that the learned user/item representation in existing MTL models can be expressed as:
\begin{equation}
\begin{aligned}
\mathbf{x}_{u}^{*} = g^{u}(\mathbf{x}_{u}, \mathbf{A}), \mathbf{y}_{i}^{*} = g^{i}(\mathbf{y}_{i}, \mathbf{A})
\end{aligned}
\end{equation}
where  $g^{u}(\cdot)$ and $g^{i}(\cdot)$ denote the representation learning function, 
$\mathbf{x}_{u}$ and $\mathbf{y}_{i}$ are the initial embeddings for user $u$ and and item $i$, and $\mathbf{A}$ is the corresponding adjacency matrix of MBG $\mathcal{G}$.
Notice that $\mathbf{A}$ is optional for $g^{u}(\cdot)$ and $g^{i}(\cdot)$ to generalize them to non-graph functions.

Taking $(\mathbf{x}_{u}^{*},\mathbf{y}_{i}^{*})$ as \textbf{\textsl{same input}} for MTL, the loss function can be formulated as:
\begin{equation}
\begin{aligned}
\mathcal{L}_{u,i}
&=\sum_{k=1}^{K}L(\hat{o}_{u,i}^{k}-{o}_{u,i}^{k})\\
&=\sum_{k=1}^{K}L(f_{k}(\mathbf{x}_{u}^{*},\mathbf{y}_{i}^{*})-{o}_{u,i}^{k})
\end{aligned}
\end{equation}
where $\hat{o}_{u,i}^{k}$ denotes the predictive probability that user $u$ will interact with item $i$ under the \textsl{k}-th behavior, ${o}_{u,i}^{k}$ is the true label, $L(\cdot)$ is the loss function, and $f_{k}(\cdot)$ is the predictive function in MTL models.
Then we have:
\begin{equation}
\begin{aligned}
{\partial{\mathcal{L}_{u,i}}\over{\partial{(\mathbf{x}_{u}^{*} \circ \mathbf{y}_{i}^{*})}}}
&=
\sum_{k=1}^{K}{\partial{L(f_{k}(\mathbf{x}_{u}^{*},\mathbf{y}_{i}^{*})-{o}_{u,i}^{k})}
\over{\partial{(\mathbf{x}_{u}^{*} \circ \mathbf{y}_{i}^{*})}}}\\
&=
\sum_{k=1}^{K}{\partial{f_{k}(\mathbf{x}_{u}^{*},\mathbf{y}_{i}^{*})}\over{\partial{(\mathbf{x}_{u}^{*} \circ \mathbf{y}_{i}^{*})}}}*{L^{'}(f_{k}(\mathbf{x}_{u}^{*},\mathbf{y}_{i}^{*})-{o}_{u,i}^{k})}\\
&=
\sum_{k=1}^{K}{a_{u,i}^{k}{\mathbf{r}^{k}}}\\
&=
\sum_{k=1}^{K}{\mathbf{r}^{' k}}
\end{aligned}
\end{equation}
where ($\circ$) is the hadamard product operation, $a_{u,i}^{k}=L^{'}(f_{k}(\mathbf{x}_{u}^{*},\mathbf{y}_{i}^{*})-{o}_{u,i}^{k})$ is a scalar. $\mathbf{r}^{k}={\partial{f_{k}(\mathbf{x}_{u}^{*},\mathbf{y}_{i}^{*})}\over{\partial{(\mathbf{x}_{u}^{*} \circ \mathbf{y}_{i}^{*})}}}$.
As $\mathbf{r}^{k}$ denotes the derivative of a scalar to a vector, it is also a vector.
$\forall$ $k \in \{1,2, \ldots, K\}$, $\mathbf{r}^{' k}$ determines the updating magnitude and direction of the vector $\mathbf{x}_{u}^{*} \circ \mathbf{y}_{i}^{*}$.
We can see that the gradients from all behaviors are coupled.
Similar to Section \ref{limitations}, we can find that there are gradient conflicts due to the coupled gradient issue if we use \textbf{\textsl{same
input}} for MTL.


\subsection{Decoupled Gradient of MESI for MTL}\label{decoupled_gradient_of_mesi}
In contrast, our proposed MESI takes \textbf{\textsl{separate
inputs}} $\mathbf{x}_{u,k}^{*}$ and $\mathbf{y}_{i,k}^{*}(k \in \{1,2, \ldots, K\})$ for MTL.
The loss function for MESI can be formulated as:
\begin{equation}
\begin{aligned}
\mathcal{L^{*}}_{u,i}
&=\sum_{k=1}^{K}L^{*}(\hat{o}_{u,i}^{k}-{o}_{u,i}^{k})\\
&=\sum_{k=1}^{K}L^{*}(h^{k}(\sum_{j=1}^{K} {\mathbf{g}_{u,i}^{k}(j) \cdot \mathbf{f}_{u,i}^{j}})-{o}_{u,i}^{k})
\end{aligned}
\end{equation}
where 
\begin{equation}
\begin{aligned}
\mathbf{g}_{u,i}^{k}=Softmax(\mathbf{W}_g(\mathbf{x}_{u,k}^{*}||\mathbf{y}_{i,k}^{*}) + \mathbf{b}_g)
\end{aligned}
\end{equation}
\begin{equation}
\begin{aligned}
\mathbf{f}_{u,i}^{k} = \mathbf{x}_{u,k}^{*} \circ \mathbf{y}_{i,k}^{*}
\end{aligned}
\end{equation}
$\hat{o}_{u,i}^{k}$ denotes the predictive probability that user $u$ will interact with item $i$ under the \textsl{k}-th behavior, ${o}_{u,i}^{k}$ is the true label, $L^{*}(\cdot)$ is the loss function used for optimization. 
And $\mathbf{g}_{u,i}^{k}$ denotes the gate for task $k$, $\mathbf{f}_{u,i}^{k}$ denotes the expert generated from input $\mathbf{x}_{u,k}^{*}$ and $\mathbf{y}_{i,k}^{*}$, which can be referred to Section \ref{mesi}. 


For arbitrary reference input vector $\mathbf{x}_{u,t}^{*}$ and $\mathbf{y}_{i,t}^{*}(t \in \{1,2, \ldots, K\})$ to be optimized, we then have:
\begin{equation}
\begin{aligned}
{\partial{\mathcal{L^{*}}_{u,i}}\over{\partial{(\mathbf{x}_{u,t}^{*} \circ \mathbf{y}_{i,t}^{*})}}}
&=
\sum_{k=1}^{K}{\partial{L^{*}(h^{k}(\sum\limits_{j=1}^{K} {\mathbf{g}_{u,i}^{k}(j) \cdot \mathbf{f}_{u,i}^{j}})-{o}_{u,i}^{k})}
\over{\partial{(\mathbf{x}_{u,t}^{*} \circ \mathbf{y}_{i,t}^{*})}}}\\
&=
\sum_{k=1}^{K}{\partial{(\sum\limits_{j=1}^{K} {\mathbf{g}_{u,i}^{k}(j) \cdot \mathbf{f}_{u,i}^{j}})}
\over{\partial{(\mathbf{x}_{u,t}^{*} \circ \mathbf{y}_{i,t}^{*})}}}*{a_{u,i}^{k}}\\
&=
\sum_{k=1}^{K}({
\partial{({\mathbf{g}_{u,i}^{k}(t) \cdot (\mathbf{x}_{u,t}^{*} \circ \mathbf{y}_{i,t}^{*})})}
\over{\partial{(\mathbf{x}_{u,t}^{*} \circ \mathbf{y}_{i,t}^{*})}}}
+{
\partial{(\sum\limits_{j=1 \atop j \neq t}^{K} {\mathbf{g}_{u,i}^{k}(j) \cdot \mathbf{f}_{u,i}^{j}})}
\over{\partial{(\mathbf{x}_{u,t}^{*} \circ \mathbf{y}_{i,t}^{*})}}}
)
*{a_{u,i}^{k}}\\
&=
\sum_{k=1}^{K}{
\mathbf{g}_{u,i}^{k}(t)}
*{a_{u,i}^{k}}
+
{
\partial{(\sum\limits_{j=1 \atop j \neq t}^{K} {\mathbf{g}_{u,i}^{t}(j) \cdot \mathbf{f}_{u,i}^{j}})}
\over{\partial{(\mathbf{x}_{u,t}^{*} \circ \mathbf{y}_{i,t}^{*})}}}
*{a_{u,i}^{t}}
+\\
&\sum\limits_{k=1 \atop k \neq t}^{K}{
\partial{(\sum\limits_{j=1 \atop j \neq t}^{K} {\mathbf{g}_{u,i}^{k}(j) \cdot \mathbf{f}_{u,i}^{j}})}
\over{\partial{(\mathbf{x}_{u,t}^{*} \circ \mathbf{y}_{i,t}^{*})}}}
*{a_{u,i}^{k}}\\
&=
\sum_{k=1}^{K}{
\mathbf{g}_{u,i}^{k}(t)}
*{a_{u,i}^{k}}
+
{
\partial{(\sum\limits_{j=1 \atop j \neq t}^{K} {\mathbf{g}_{u,i}^{t}(j) \cdot \mathbf{f}_{u,i}^{j}})}
\over{\partial{(\mathbf{x}_{u,t}^{*} \circ \mathbf{y}_{i,t}^{*})}}}
*{a_{u,i}^{t}}
+0\\
&=
\sum_{k=1}^{K}{{
\mathbf{g}_{u,i}^{k}(t)}
*{a_{u,i}^{k}}}
+
{
\sum\limits_{j=1 \atop j \neq t}^{K}{\partial{({\mathbf{g}_{u,i}^{t}(j) \cdot \mathbf{f}_{u,i}^{j}})}
\over{\partial{(\mathbf{x}_{u,t}^{*} \circ \mathbf{y}_{i,t}^{*})}}}
*{a_{u,i}^{t}}}
\end{aligned}
\end{equation}
where 
$$
{a_{u,i}^{k}} = {({h^{k}}^{'}(\sum_{j=1}^{K} {\mathbf{g}_{u,i}^{k}(j) \cdot \mathbf{f}_{u,i}^{j}})}
*{{L^{*}}^{'}(h^{k}(\sum_{j=1}^{K} {\mathbf{g}_{u,i}^{k}(j) \cdot \mathbf{f}_{u,i}^{j}})-{o}_{u,i}^{k})}
$$
is a scalar.

In the above derivation process, it can be clearly seen that our proposed MESI decouples the gradients of different behaviors and selectively uses information of different behaviors to update the gradients, which alleviates the issue of gradient conflict.

\subsection{Calculation of Pearson Correlation}
\label{Calculation_of_Pearson_correlation}
We choose the Pearson correlation to divide users into different test groups.
The Pearson correlation between behavior $i$ and $j$ for user $u$ can be calculated as follows:
\begin{equation}
r_u^{s,t}=\frac{\sum\limits_{j=1}^{N}\left(\mathbf{Y}_{u,j}^s-\bar{\mathbf{Y}}_u^s\right)\left(\mathbf{Y}_{u,j}^t-\bar{\mathbf{Y}}_u^t\right)}{\sqrt{\sum\limits_{j=1}^{N}\left(\mathbf{Y}_{u,j}^s-\bar{\mathbf{Y}}_u^s\right)^{2}} \sqrt{\sum\limits_{j=1}^{N}\left(\mathbf{Y}_{u,j}^t-\bar{\mathbf{Y}}_u^t\right)^{2}}}
\end{equation}
where $\mathbf{Y}_{u,j}^s$ and $\mathbf{Y}_{u,j}^t$ denote the entries at the $u$-th row and $j$-th column of user-item interaction matrices $\mathbf{Y}^s$ and $\mathbf{Y}^t$ respectively,
$\bar{\mathbf{Y}}_u^s$ and $\bar{\mathbf{Y}}_u^t$ denote the mean of the input vector $\mathbf{Y}_u^s$ and $\mathbf{Y}_u^t$.
$N$ is the length of the input vector, which is also the number of items.
After we have obtained the Pearson correlation between each pair of behaviors, we can get the final average Pearson correlation among all behaviors for each user $u$ as:
\begin{equation}
r_u= \frac{2}{K(K-1)}\sum\limits_{s=1}^{K-1}\sum\limits_{t=s+1}^{K}{r_u^{s,t}}
\end{equation}

\subsection{Analysis of Label Correlations}\label{Label_Correlations}
\begin{figure}[h]
	\centering
	\setlength{\belowcaptionskip}{-0.3cm}
	\setlength{\abovecaptionskip}{0cm}
	\includegraphics[width=0.35\textwidth]{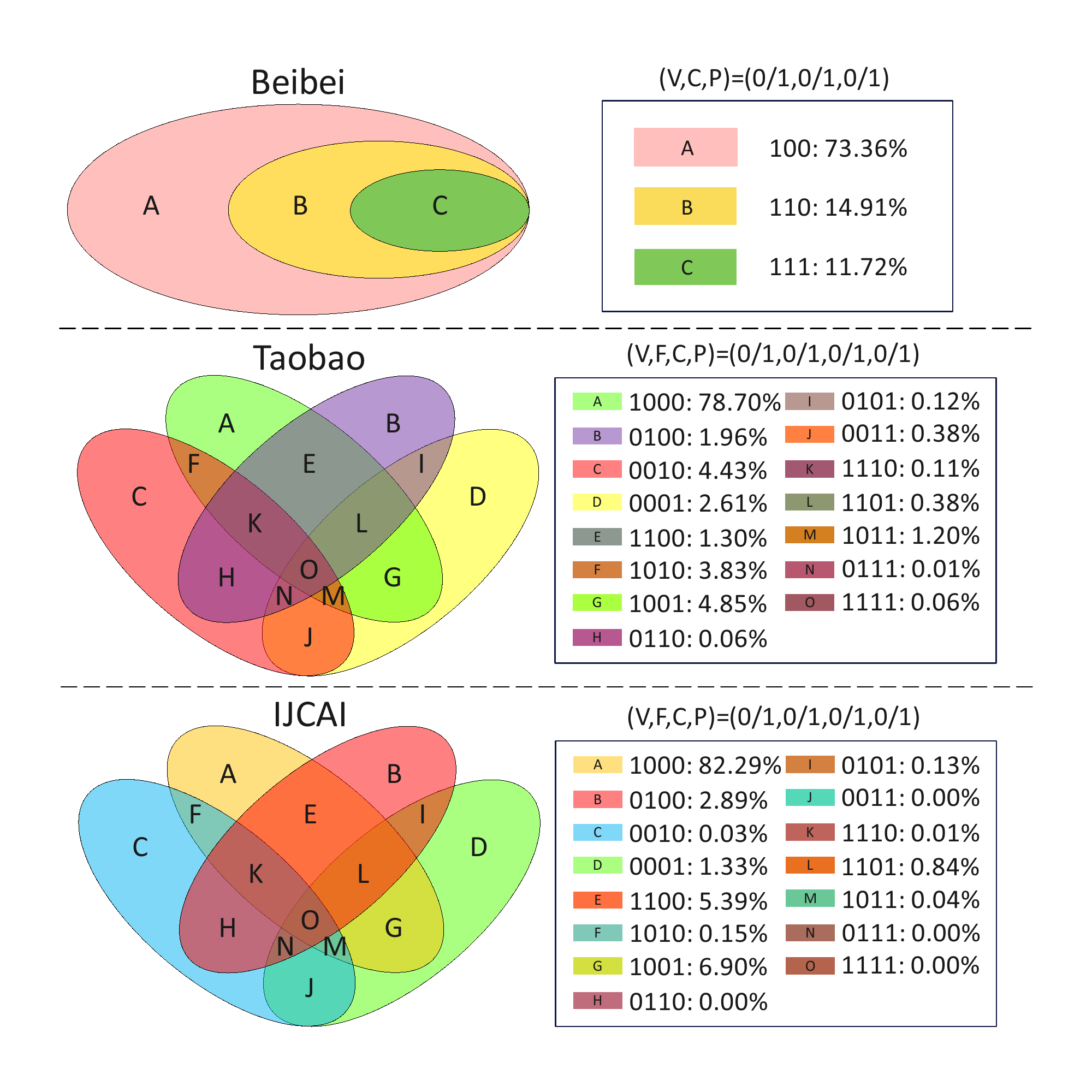}
    \Description{Venn diagram of label correlations on the three datasets. 1/0 means have or not have this type of behavior. E.g., 0110 represents those users who only have favorite and cart behaviors with items.}
	\caption{Venn diagram of label correlations on the three datasets. 1/0 means have or not have this type of behavior. E.g., 0110 represents those users who only have favorite and cart behaviors with items.}
	\label{fig:statistics}
\end{figure}
\label{Label_Correlations}
The multi-behavior data can be treated ``\textsl{as labels}'' for multi-task supervised learning.
Figure \ref{fig:statistics} shows the label correlations with the venn diagram when treating multi-behavior data as labels, where different overlaps represent different label correlations.

\subsection{Analysis of Complexity and Efficiency}
\label{complexity and efficiency}
\subsubsection{Complexity Analysis}
\textbf{Time Complexity.}
We analyze the time complexity of CIGF where the CIGCN module is the main cost.
The computational complexity for CIGCN is $\sum_{k=1}^{K}\sum_{l=1}^{L}{O\left(\left|\mathcal{B}_{v,k}^l \right| \cdot d\right)}$, where $\left|\mathcal{B}_{v,k}^l \right|$ denotes the number of edges existed in all graphs of set $\mathcal{B}_{v,k}^l$, $K$ is the behavior number, $L$ is the layer number and $d$ is the embedding size.
In CIGCN, the dense graphs $\mathbf{B}_{v,k}^{l,s}$ in set $\mathcal{B}_{v,k}^l$ are transformed into $l$ sparse graph for computation. 
As $l$ is usually very small, the time complexity is comparable with existing GNNs, which is further verified with experiments in Section \ref{efficiency_comparison}.
\newline
\textbf{Space Complexity.}
The learnable parameters in our proposed CIGF are mainly from the user and item embedding $\mathbf{x}_u$ and $\mathbf{y}_i$, which is similar to existing GNNs.
Besides, as dense graph $\mathbf{B}_{v,k}^{l,s}$ in set $\mathcal{B}_{v,k}^l$ are transformed into sparse behavior-specified graphs $\mathcal{G}^1, \mathcal{G}^2, \cdots, \mathcal{G}^K$ for computation, no additional memory space is needed to store these graphs, which makes the memory footprint of the intermediate process acceptable.

\subsubsection{Efficiency Analysis}
\label{efficiency_comparison}
\begin{table}[H]
\setlength{\abovecaptionskip}{0cm}
\setlength{\belowcaptionskip}{-0.1cm}
\caption{Training time comparison (seconds per epoch) of different methods on all three datasets.}
    \centering
\setlength{\tabcolsep}{1mm}{
\small
\begin{tabular}{c|c|c|c}
\midrule[0.25ex]
\diagbox{Model}{Training time (s)}{Dataset} &
\multicolumn{1}{c|}{Beibei} &
\multicolumn{1}{c|}{Taobao} & 
\multicolumn{1}{c}{IJCAI} \\ \hline 
GHCF & \textbf{8.31} & 20.02 & -      \\
MB-GMN & 14.95 & 27.03 & 79.25     \\
CIGF & 10.37 & \textbf{16.78} & \textbf{61.60}     \\
\hline \hline
\end{tabular}}
\label{tab:Time complexity}
\vspace{-2mm}
\end{table}

\begin{table}[H]
\setlength{\abovecaptionskip}{0cm}
\setlength{\belowcaptionskip}{-0.1cm}
\caption{Testing time comparison (seconds per epoch) of different methods on all three datasets.}
    \centering
\setlength{\tabcolsep}{1mm}{
\small
\begin{tabular}{c|c|c|c}
\midrule[0.25ex]
\diagbox{Model}{Testing time (s)}{Dataset} &
\multicolumn{1}{c|}{Beibei} &
\multicolumn{1}{c|}{Taobao} & 
\multicolumn{1}{c}{IJCAI} \\ \hline 
GHCF & 9.88 & 34.48 & -      \\
MB-GMN & 2.96 & 19.83 & 68.77     \\
CIGF & \textbf{2.46} & \textbf{18.79} & \textbf{59.56}     \\
\hline \hline
\end{tabular}}
\label{tab:Test Time complexity}
\vspace{-2mm}
\end{table}

Apart from the model effectiveness, the training efficiency also matters.
Table \ref{tab:Time complexity} shows the training time (one epoch) comparison between our CIGF and two representative baselines on all three datasets.
The best baseline MB-GMN requires the longest training time, while our CIGF is faster with \textbf{30.64\%}, \textbf{37.92\%}, and \textbf{22.27\%} time reduction on the three datasets.
Besides, though GHCF is slightly faster than our CIGF on the Beibei dataset, it inapplicable to the IJCAI dataset due to the unaffordable memory usage brought by non-sampling learning. Besides, as shown in Table \ref{tab:Test Time complexity}, our proposed CIGF is \textbf{16.89\%}, \textbf{5.24\%}, and \textbf{13.39\%} faster than the fastest of the other models on three datasets. GHCF performs well in training efficiency, while it performs worst in testing. The possible reason is that the non-sampling learning loss of GHCF dramatically improves the efficiency of loss calculation, thus significantly improving training efficiency. While the GNN part, which contributes to the main complexity of GHCF, is more complicated, so it takes more time to test.
In summary, we claim that CIGF has the best overall efficiency.